\documentclass[useAMS,usenatbib]{mn2e}
\usepackage{graphicx,fleqn,times,color}
\usepackage{lscape}
\usepackage{hyperref}

\title[Bars in galaxies with gas and a triaxial halo]{Bar formation
  and evolution in disc galaxies with gas and a triaxial halo:
  Morphology, bar strength and halo properties}

\author[E. Athanassoula, R. Machado and S. Rodionov
]{E. Athanassoula$^{1}$\thanks{E-mail:
lia@oamp.fr}, Rubens E. G.
Machado$^{1,2}$, S.A.~Rodionov$^{1,3}$\\
$^{1}$Aix Marseille Universit\'e, CNRS, LAM (Laboratoire
d'Astrophysique de Marseille)      
UMR 7326, 13388, Marseille, France\\
$^{2}$Instituto de Astronomia, Geof\'isica e Ci\^encias Atmosf\'ericas,
Universidade de S\~ao Paulo,
R. do Mat\~ao 1226, 05508-090 S\~ao Paulo, Brazil\\
$^{3}$Sobolev Astronomical Institute,
St. Petersburg State University,
Universitetskij pr.~28, 198504 St. Petersburg, Stary Peterhof, Russia\\
}

\begin{document}

\date{Accepted ???? ??? ??. Received ???? ??? ??}

\pagerange{\pageref{firstpage}--\pageref{lastpage}} \pubyear{2012}

\maketitle

\label{firstpage}

\begin{abstract}

We follow the formation and evolution of bars in N-body simulations of
disc galaxies with gas and/or a triaxial halo. We find that both the
relative gas fraction and the halo shape play a major role in the
formation and evolution of the bar. In gas-rich simulations, the disc 
stays near-axisymmetric much
longer than in gas-poor ones, and, when the bar starts growing, it
does so at a much slower rate. Due to these two effects combined,
large-scale bars form much later in gas-rich than in gas-poor discs. This can
explain the observation that bars are in place earlier in massive red
disc galaxies than in blue spirals. We also find that the
morphological characteristics in the bar region are strongly influenced 
by the gas fraction. In particular, the bar at the end of the
simulation is much weaker in gas-rich cases. The quality 
of our simulations is such as to allow us to discuss the question of
bar longevity because the resonances are well resolved and the number
of gas particles is sufficient to describe the gas flow adequately. In no case
did we find a bar which was destroyed. 

Halo triaxiality has a dual
influence on bar strength. In the very early stages of the simulation
it induces bar formation to start earlier. On the other hand, during the
later, secular evolution phase, triaxial haloes lead 
to considerably less increase of the bar strength than spherical ones.
The shape of the halo evolves considerably with
time. We confirm previous results of gas-less simulations that
  find that the inner part of an initially spherical halo can become elongated and
develop a halo bar. However we also show that, on the contrary, in
gas rich simulations, the inner parts of an initially triaxial halo
can become rounder with time. The
main body of initially triaxial 
haloes evolves towards sphericity, but in initially strongly triaxial
cases it stops well short of becoming spherical. Part of the angular
momentum absorbed by the halo generates considerable rotation of the
halo particles that stay located relatively near the disc for long
periods of time. Another part generates halo bulk
rotation, which, contrary to that of the bar, increases with time but stays
small. Thus, in our models there are two non-axisymmetric components
rotating with different pattern speeds, namely the halo and the bar,
so that the resulting dynamics have strong similarities to the
dynamics of double bar systems. 

\end{abstract}



\section{Introduction}
\label{sec:intro}

Bars, weak or strong, are present in the majority of present-day disc galaxies
\citep[e.g.][]{deVaucouleurs.P.91, Eskridge.P.00, Knapen.Shlosman.Peletier.00,
Whyte.AMEFP.02, Marinova.Jogee.07, Menendez.SSJS.07, Barazza.JM.08,
Aguerri.MC.09, Marinova.P.09, Mendez.SA.10, Masters.P.2011}, with an often
quoted fraction of roughly two thirds. They can also be found at higher
redshifts \citep{Abraham.TSEGB.96, vandenBergh.AETSG.96, Abraham.METB.99,
Elmegreen.EH.04, Jogee.P.04}, although there they constitute a smaller fraction
of the disc galaxies than at low redshifts \citep{Sheth.P.08, Nair.Abraham.10}. 
A number of observational studies, ranging from in-depth studies of
single objects to large surveys, have provided useful information on
the morphological, photometrical and kinematical properties of bars
\citep[e.g.][]{Sheth.VRTT.05, Gadotti.ACBSR.07, Marinova.P.09,
 Buta.LSK.10, Gadotti.11, Hoyle.P.11, Laurikainen.SBK.11, Masters.P.2011,
 Simard.MPEM.11, Beirao.P.12, Martinez-Garcia.12,
  Perez.AMA.12,Wang.P.12}.

This observational effort was accompanied by a considerable effort
with N-body simulations  (e.g. \citealt{Debattista.Sellwood.00};
  \citealt{Athanassoula.Misiriotis.02}; \citealt{Athanassoula.02};
  \citealt{Athanassoula.03}, hereafter A03;
  \citealt{ONeil.Dubinski.03}; \citealt{Valenzuela.Klypin.03}; \citealt{Martinez.VSH.06};
  \citealt{HolleyBockelmann.WK.05}; \citealt{Dubinski.BS.09}). Such simulations 
provided information on the formation and evolution of barred galaxies, on the role
of the halo and on the redistribution of angular
momentum within the galaxy. They furthermore
allowed detailed comparisons with 
observations. Even so, this simulation work can be considered as the first
step, since it has, by necessity, relied on a number of simplifying
approximations. In this paper, we will revisit two such
approximations, and consider bar formation and evolution in their
absence, i.e. in more realistic cases than in previous studies.

One approximation used in the vast majority of previous studies is that
initially the halo is spherically symmetric. Yet cosmological
simulations \citep[e.g.][]{Dubinski.Carlberg.91, Jing.Suto.02,
Bailin.Steinmetz.05, Allgood.FPKWFB.06, Novak.P.06, 
Vera-Ciro.SHFNSVW.11, Schneider.FC.12} 
have shown that in cases with no baryons the haloes are strongly
triaxial, as could be expected by the fact that the halo shapes 
can be strongly modified during interactions and mergings
\citep{McMillan.AD.07, Kazantzidis.P.04}, as well as by the radial orbit
instability
\citep{Merritt.Aguilar.85, Barnes.HG.86, Dejonghe.Merritt.88,
Aguilar.Merritt.90, Weinberg.91, Cannizzo.Hollister.92,
Huss.JS.99, Boily.Athanassoula.06, MacMillan.P.06, Bellovary.P.08} 

Yet observations show that present day haloes should be considerably
more axisymmetric in the equatorial plane than the above mentioned
papers suggest \citep[e.g.][]{Trachternach.P.08}, while haloes in
cosmological simulations including 
baryons are less triaxial. It is thus necessary to
understand the effect of baryons on the evolution of halo shapes in barred 
galaxies. Aspects of this question have been already addressed in several
papers \citep[][hereafter MA10]{Dubinski.94, Gadotti.deSouza.03, 
  Curir.MM.06, Berentzen.Shlosman.Jogee.06,
  Berentzen.Shlosman.06, Athanassoula.07, Widrow.08, Debattista.P.08,
  Machado.Athanassoula.10}, but a full understanding is still not available. 

A second, often used approximation consists in either neglecting the gas
component, or modelling it in an oversimplified way. Yet gas has a
considerable effect on the evolution of disc galaxies. Its mass may be
a small fraction of the total at present, but it has been much more
important in the past \citep[e.g.][]{TacconiP.10}. Furthermore,
gas, being a cold component, can respond quite strongly to
gravitational perturbations. A number 
of studies including gas have given important insight
\citep[e.g.][]{Berentzen.HSF.98, Bournaud.Combes.02, Berentzen.AHF.03,
  Berentzen.AHF.04, Bournaud.CS.05, Debattista.MCMWQ.06, 
  Berentzen.SMVH.07, Heller.SA.07L, Heller.SA.07,
  Wozniak.Dansac.09, Villa.VSH.10, DeBuhr.MW.12}, but
relatively few had a sufficient number of particles
\citep{Patsis.Athanassoula.00}, and most of them
neglected the physics of the gas, i.e. neglected its
multi-phase nature, as well as the related star
formation, feedback and cooling. In these simplified
cases the amount of gas stays constant during the simulation, so, if
the simulation spans several Gyr, the adopted gas fraction
is too low during the first part of the simulation and/or too high
during the last part.

Since this paper was submitted, two new papers on related subjects
were published. They both use cosmological zoom re-simulations 
and include gas with a realistic physics. The first
one \citep{Kraljic.BM.12} measures the fraction 
of disc galaxies that are barred and compares them to observations,
while the second one \citep{Scannapieco.Athanassoula.12} studies bar
properties of two bars in Aquarius galaxies. 

In the present paper we follow the formation and evolution of a bar
in a disc galaxy with a triaxial halo and which includes a gaseous
disc component undergoing star formation, feedback
and cooling. In Sect.~\ref{sec:techniques} we give information on the
numerical aspects of the work. In particular, we describe our use of
the \textsc{gadget2} code, how the
equilibrium initial conditions were derived and what their relevant
properties are (Sect.~\ref{subsec:initialconditions}). Results are 
given and discussed in Sects.~\ref{sec:globalevol} to ~\ref{sec:oscillations}. 
Sect.~\ref{sec:globalevol} gives the evolution with time of the gas fraction
and of the general morphology. In
Sect.~\ref{sec:barstrength} we discuss the evolution of the bar
strength with time and in Sect.~\ref{sec:longshortlived} we enter the
debate of whether bars are long-lived or short-lived. 
In Sect.~\ref{sec:haloprop} we present the radial profiles and the time
evolution of the halo axial ratios, as well as the kinematics of the
halo material. The latter together with the angular momentum
redistribution within the galaxy we relate to the bar
strength. Interaction between the  various non-axisymmetric
  components is the subject of Sect.~\ref{sec:oscillations}. We
present further discussion of our results in Sect.~\ref{sec:discussion}
and conclude in Sect.~\ref{sec:conclusions}. 

\section{Techniques}
\label{sec:techniques}

\subsection{Simulations}
\label{subsec:simu}

We use a version of \textsc{gadget2} including gas and its physics
\citep{Springel.YW.2001, Springel.Hernquist.02, Springel.05}. The dark  
matter and the stars are followed by N-body particles and gravity is
calculated with a tree code. The code uses an improved SPH
method \citep{Springel.Hernquist.02} and sub-grid
physics \citep{Springel.Hernquist.03}. In this approach, each
SPH particle represents a region of the ISM containing both cold
gas clouds and hot ambient gas, the two in pressure equilibrium.   

In the following we use the \textsc{gadget2} system of units; i.e. the
unit of length is 1 kpc, the unit of mass is 10$^{10}$ $M_\odot$ and
the unit of velocity 1 km/sec. As a result the time unit is 0.98 Gyr.
For simplicity, and given the accuracy of all our measurements, we
used a time unit of one Gyr. We continued all simulations up to 10 
Gyr. This is longer than what is expected for the 
combined bar formation and evolution phases in real disc galaxies,
but allows us to follow fully all secular evolution phases. For
comparisons with observed nearby galaxies, however, times between 6 and 8 Gyr
may be more appropriate. For this reason we will include in many of
our discussions information from both the 6 and the 10 Gyr results.   
We used a softening length of 50 pc for all components and an opening
angle for the tree-code of 0.5. 

\textsc{gadget2} offers the possibility of using several types of
particles for the various components of the galaxy. In the following we
will use four types, namely: \textsc{halo}, \textsc{disk},
\textsc{gas} and \textsc{stars}. The \textsc{disk} particles represent stars
already present in the initial conditions and their number 
remains constant throughout the simulation. But as the simulation
evolves, gas particles give rise to new stars, via star
formation, so that both the \textsc{disk} particles and the 
\textsc{star} particles represent the stars of the galaxy. To
distinguish between the different components, we will use in the
following the words \textsc{disk} and \textsc{stars} when we talk
about the corresponding \textsc{gadget2} components, the words `stellar
disc' when we refer to all stellar particles in the galactic disc
(i.e. when we refer to both the \textsc{disk} and \textsc{stars}
particles together) and the word `disc' when we refer to all 
components in the disc, 
i.e. \textsc{disk}, \textsc{stars} and \textsc{gas}. The
\textsc{disk} component may be thought of as 
representing an older stellar population, whereas the \textsc{stars}
particles represent a mix of stars ranging in age from very young to
as old as the simulation time. The gas and halo particles, where no
confusion is possible, will be simply referred to as `gas' and `halo',
respectively.  

\subsection{Initial conditions}
\label{subsec:initialconditions} 

\begin{figure}
\includegraphics[scale=0.45]{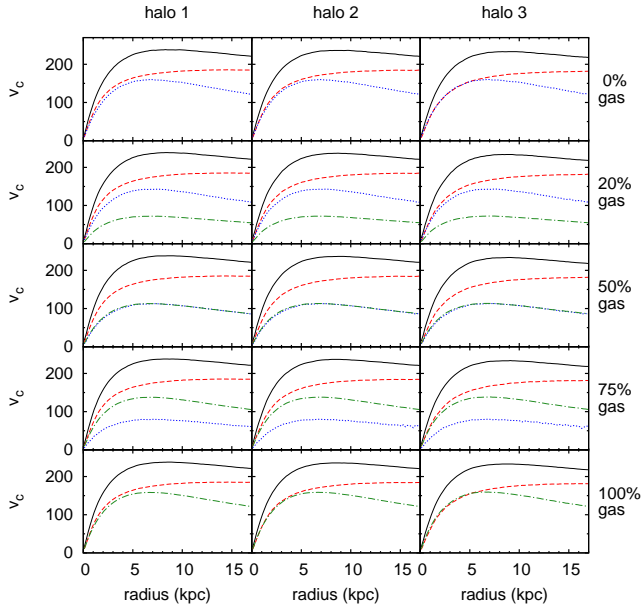}
\caption{Circular velocity curves of the initial
  conditions ($t$ = 0): total (solid
  lines, black in the online version), halo (dashed, red on the online
  version), disk (dotted, blue in the online version) and gas
  (dot-dashed, green in the online version). The three columns,from
  left to right, 
  correspond to the three haloes used in this paper, namely halo 1,
  halo 2 and halo 3. Each row corresponds to a different initial
  gas fraction (calculated as the fraction of gas in the disc component).
  } 
\label{fig:vcirc}
\end{figure}

Information on the initial conditions of all models used in this paper
is given in Table~\ref{tab:allmodels}. The first four columns give the
run number, the halo which is used and its $b/a$ and $c/a$ axial
ratios, respectively. The semi-major axes $a$, $b$ and $c$ are placed
along the $x$, $y$ and $z$ axes respectively. The fifth column gives
the fraction of gas in the disc component and the sixth one 
gives the number of particles in the gas component.

The initial conditions have been built using the iterative method
\citep*{Rodionov.AS.09}, and, more specifically, its 
extension to include a gaseous component
\citep{Rodionov.Athanassoula.11}. We make a series of short constrained 
iterative steps to build each component in near-equilibrium in the 
total galactic potential. In this way we avoid transients which could
affect the instabilities under study. In the triaxial models, 
both the stellar and the gaseous disc have an elliptical shape (see
\citealt{Rodionov.AS.09} for details). 
The circular velocity curves of the initial conditions are
shown in Fig.~\ref{fig:vcirc}. They are essentially the
same for all models, as required for a study of the effect of gas
fraction and halo triaxiality.
  
It should be stressed that a simulation from the present work can not
be directly compared with a simulation from MA10 with the same halo initial
axial ratios, because of the difference in how the initial
conditions were made. Indeed in MA10 we first prepared an equilibrium
halo model with the desired axial ratio. Then a disc was grown in this
halo, which brings about an axisymmetrisation of the latter component
of the order of
half its final axisymmetrisation. So the halo potential that the bar
feels as it grows is considerably less triaxial than that of the
initial halo model. For example, when, in MA10, we started with a
halo model of 
axial ratios $b/a$ = 0.8 and $c/a$ = 0.6, we obtained after the
disc was introduced a rounder halo with an axial ratio $b/a$ of 
$\sim0.9$. This is not the case here, where the disc is 
built in equilibrium within the halo having the prescribed axial
ratio. So if we want to compare one of our simulations with 
that particular MA10 simulation, we would have to use one with $b/a$ =
0.9 and not 0.8. Further differences concern the disc, whose shape
here is obtained by the iterative method in its search for
near-equilibrium.  

The initial azimuthally averaged density distribution of the stellar disc is given by

\begin{equation}
\rho_d (R, z) = \frac {M_d}{4 \pi h^2 z_0}~~exp (- R/h)~~sech^2 (\frac{z}{z_0}),
\end{equation}

\noindent
where $R$ is the cylindrical radius, $M_d$ is the disc mass, $h$ is the
disc radial scale length and $z_0$ is the disc vertical scale thickness. 
The radial scale length is $h=3$~kpc and the scale height is
$z_0=0.6$~kpc. For the gas we adopt the same radial profile and the
same scale length. This is necessary in order to be able to make
sequences of models where only the gas fraction changes and all the
remaining parameters and quantities are the same. The gas scale height
is considerably smaller than that of the stars and its precise value is
set by the hydrostatic equilibrium achieved during the iterative
calculation of the initial conditions \citep{Rodionov.Athanassoula.11}.
The total disc mass (stellar plus gaseous) is always $M_{d}=5 \times
10^{10}~M_{\odot}$.  

When creating the initial conditions we impose a radial velocity
dispersion for the \textsc{disk} particles, $\sigma_R(R)$, of

\begin{equation}
\label{eq_svR}
\sigma_R(R) = 100 \cdot \exp\left(-R/3h\right) \; {\rm km \, s^{-1}} \, . 
\end{equation}

All haloes have been built so as to have, within the allowed accuracy,
the same spherically averaged initial radial profile, namely
 
\begin{equation}
\rho_h (r) = \frac {M_h}{2\pi^{3/2}}~~ \frac{\alpha}{r_c} ~~\frac
{exp(-r^2/r_c^2)}{r^2+\gamma^2},
\end{equation}

\noindent
where $r$ is the radius, $M_h$ is the mass of the halo and $\gamma$
and $r_c$ are the halo core and cut-off radii. The parameter $\alpha$
is a normalisation constant defined by 

\begin{equation}
\alpha = [1 - \sqrt \pi~~exp (q^2)~~(1 -erf (q))]^{-1},
\end{equation}

\noindent
where $q=\gamma / r_c$ \citep[][]{Hernquist.93}. In all simulations we
take $\gamma=1.5$~kpc, $r_{c}=30$~kpc and $M_{h}=2.5 \times 10^{11}~M_{\odot}$.

This model has several advantages. Compared to observations, its
  rotation curve (Fig. \ref{fig:vcirc}) has a realistic shape. We have
  also avoided a strong cusp in the centre, in good agreement with
  observations \citep[e.g.][]{deBlok.MBR.01, deBlok.Bosma.02,
    deBlok.BM.03, Simon.BLB.03, Kuzio.MBB.06, 
Kuzio.MB.08, deBlok.WBTOK.08, Oh.BWBK.08, Battaglia.ITHHLJ.08,
 deBlok.10, Walker.P.11, Amorisco.Evans.12,
 Penarrubia.PWK.12}. Finally, this model has been used in a number of
  previous studies, on which we were able to rely here. For example, 
  \cite{Athanassoula.02} determined the number of particles which we
  need to have in a halo of this type in order to describe its
  resonances adequately for the evolution. This necessitated the
  calculation of a large number of orbits in order to determine when they
  get kicked in or out of resonance by particle noise. As this was
  done for a range of softening values (Athanassoula, unpublished), we
  were able to rely here on these results. 

At the request of the referee, we compare here quantitatively our halo
profile with that of a cosmologically motivated NFW profile
\citep{Navarro.FW.96}. The main difference of course is that the NFW
profile has a cusp in the innermost parts, while ours has a core. As
already mentioned above, we made this choice in order to be in
agreement with observations. To compare the two profiles at larger
radii, we tried fitting our profile with an NFW profile in the region one to 
20 kpc, but we could get an acceptable  fit only for large values of the
concentration parameter, $c_{200} > 30$, which are much larger than
the cosmologically motivated values. This comparison,  
however, is not fair. Indeed, in our iterative solution the disc and 
halo are built in equilibrium within each
other, while the NFW profile is a stand-alone component. For a fair comparison
we need to calculate the adiabatic contraction of a 
cosmologically motivated NFW halo ($M_{200} = 10^{12} M_{\odot}$
and $c_{200}$ = 8.07, see \citealt{Prada.KCBRP.12}), due to the disc we
adopted (eq. (3)) and compare its cumulative mass profile to that of
our halo profile. We used the method
described in \cite{Gnedin.KKN.04} and found an agreement better than,
or of the order of 10\% in the region 3 to 35 kpc, 
i.e. everywhere except the innermost region (as could be expected
because we -- wilfully -- used a core while the NFW profile has a
cusp) and the outermost region beyond 35 kpc, where our gradual
cut-off becomes important. Note, however, that moving this cutoff
outwards by a factor of three does not change the results of the
simulations significantly for spherically symmetric haloes. Thus we
can conclude that our results are not incompatible with a
cosmologically motivated NFW profile, within a radius range of more than 10
disc scale lengths, which includes the main bar resonances,
and which excludes of course the central cusp/core region. 


In all simulations presented here the halo is represented by $10^6$
particles and 
the mass of each halo particle is $m_{halo}= 2.5 \times
10^{5}~M_{\odot}$. The total disc mass is always one fifth of the
total halo mass, but the disc particles are initially distributed in
two 
components: the gas particles and the \textsc{disk} particles. Because
each set of three models has a different fraction of the total disc
mass in the form of gas, the numbers of \textsc{disk} and gas
particles are different. This is made in such a way that the mass of
each gas particle is always the same in all initial conditions:
$m_{gas}= 5 \times 10^{4}~M_{\odot}$. Likewise, $m_{\textsc{disk}}= 2.5 \times
10^{5}~M_{\odot}$ is the same in all simulations. The \textsc{disk}
and halo mass resolutions are the same, but the gas mass is more
resolved. Indeed preliminary test simulations showed that a
high number of gas particles is necessary in order to
describe reasonably well the gas component \citep[see
also][]{Patsis.Athanassoula.00}, considerably more than what
has been used in most previous such studies. To make sure that this number
of particles is sufficient we ran one simulation (simulation 111 in Table
\ref{tab:allmodels}) four more times, with one quarter, one half,
double and quadruple the number of particles for all components,
i.e. reaching two million of gas particles. We
also re-ran simulation 106 with five times more stellar and dark
matter particles and 20 times as many gas particles,
i.e. reaching 4 million gas particles. Comparing all these 
runs showed us that for the global properties discussed here, the
resolution we have adopted is well sufficient.  

\begin{table}
\caption{Properties of model initial conditions}
\label{tab:allmodels}
\begin{center}
\begin{tabular}{l c c c c c}
\hline
run & halo & $b/a$ & $c/a$ & gas & N$_{gas}$\\
&  & & & fraction & \\
\hline
 101  & 1 & 1.0 & 1.0 & 0.0 &                  0 \\
 102  & 2 & 0.8 & 0.6 & 0.0 &                  0 \\
 003 & 3 & 0.6 & 0.4 & 0.0 &                 0 \\
\hline
 106 & 1 & 1.0 & 1.0 & 0.2 & $2~ \times 10^5$ \\
 109 & 2 & 0.8 & 0.6 & 0.2 & $2~ \times 10^5$ \\
 110 & 3 & 0.6 & 0.4 & 0.2 & $2~ \times 10^5$ \\
\hline
 111 & 1 & 1.0 & 1.0 & 0.5 & $5~ \times 10^5$ \\
 114 & 2 & 0.8 & 0.6 & 0.5 & $5~ \times 10^5$ \\
 115 & 3 & 0.6 & 0.4 & 0.5 & $5~ \times 10^5$ \\
\hline
 116 & 1 & 1.0 & 1.0 & 0.75 & $7.5~ \times 10^5$ \\
 117 & 2 & 0.8 & 0.6 & 0.75 & $7.5~ \times 10^5$ \\
 118 & 3 & 0.6 & 0.4 & 0.75 & $7.5~ \times 10^5$ \\
\hline
 119 & 1 & 1.0 & 1.0 & 1. &  $1~ \times 10^6$ \\
 120 & 2 & 0.8 & 0.6 & 1. &  $1~ \times 10^6$ \\
 121 & 3 & 0.6 & 0.4 & 1. &  $1~ \times 10^6$ \\
\end{tabular}
\end{center}
\end{table}

\subsection{Analysis}
\label{subsec:analysis}

In \cite{Athanassoula.Misiriotis.02} and MA10 we described a number of
techniques for measuring quantities 
relevant to the particle distributions. Unless stated otherwise, we
will use them also here and refer the reader to the above mentioned
two papers for a full description. 

\subsubsection{Halo shape}
\label{subsubsec:haloshape}

To measure the axial ratios of the halo, we first calculate the local
halo density at the position of each halo particle using its distance
to its nearest neighbours, then
sort out these particles with respect to this local density,
divide them in bins of equal particle number and
then calculate the axial ratios from the eigenvalues of the inertia
tensor calculated within each bin. In this way we avoid the bias which
would have been introduced had we sorted the particles with respect to
distance from the centre, as initially pointed out by
\cite{Athanassoula.Misiriotis.02}. For each bin we also calculate the
mass weighted mean radius. If we wish to have information
within a given range of radii, we take the average of those bins whose
mean radius is included in the chosen range.   
 
\subsubsection{Bar strength}
\label{subsubsec:barstrength}

The strength of the bar is not a uniquely defined quantity, and in
fact any function of the bar mass, axial ratio and length can be
considered. Thus many different definitions have been so far
used. Here we will use a particularly straightforward one, based
on the Fourier components of the bi-dimensional mass distribution 
\begin{equation}
a_m (R) = \sum _{i=0}^{N_{R}}~m_{i}~\cos (m\theta_i), ~ m=0, 1, 2, ...
\end{equation}
\begin{equation}
b_m (R) = \sum _{i=0}^{N_{R}}~m_{i}~\sin (m\theta_i), ~ m=1, 2, ... 
\end{equation}
where $N_{R}$ is the number of particles inside a given annulus 
around the cylindrical radius $R$, $m_{i}$ is the
mass of particle $i$ and $\theta_i$ its azimuthal angle. The $a_m (R)$
and $b_m (R)$, however, are a function of the cylindrical radius,
while we need to characterise the bar strength by a single number at
every time and for every simulation. We
will thus measure the bar strength by the maximum
amplitude of the relative $m$ = 2 component, namely 
\begin{equation} \label{eq:Am}
A_2 = max \left(\frac{\sqrt{a_2^2+b_2^2}}{a_0}\right).
\label{eq:A2max}
\end{equation}
The cylindrical radius at which this maximum occurs will be denoted by
$R_{max}$. We verified that this measure of the bar strength gives
qualitatively similar results to those of the other measures used
e.g. in MA10 or A03, while being more
straightforward to implement for simulations with gas, where
considerable spiral structure can be present and the form of the $A_2
(R)$ curves can be considerably perturbed. 

\section{Global evolution}
\label{sec:globalevol}

\subsection{Gas fraction}
\label{subsec:gasfraction}

\begin{figure}
\includegraphics[scale=1.]{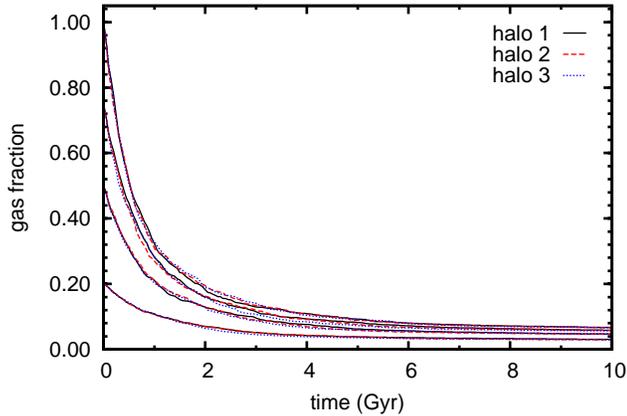}
\caption{Fraction of gas in the disc component as a function of time,
  for all simulations discussed here. We use different line styles and
  colours for simulations with different haloes, as given in the upper
  right corner. It is clear that the gas fraction at any given time
  depends strongly on the initial gas fraction, but hardly on the halo initial
  ellipticity.} 
\label{gasfraction}
\end{figure}

Fig.~\ref{gasfraction} displays the evolution of the gas fraction with
time. It shows that this decreases with time, as expected because of
the star formation. This decrease is quite steep during the first couple
of Gyr, when the gas fraction is still high, and then flattens
out. Fig.~\ref{gasfraction} also shows that the shape of the halo hardly, 
if at all, influences the total amount of stars formed and therefore the
amount of gas left at any time. 

\begin{table}
\caption{Fraction of gas in the disc for different times (in Gyr,
  left column)}
\label{tab:gasfract}
\begin{center}
\begin{tabular}{l c c c c}
\hline
time & & ~~~~~~gas fraction & &  \\
\hline
0  & 20\% & 50\% & 75\% & 100\% \\
2  & 7\% & 13\% & 16\% & 19\% \\
5  & 4\% & 6\% & 8\% & 9\% \\
10 & 3\% &  5\% & 6\% & 7\% \\

\hline
\end{tabular}
\end{center}
\end{table}

Table~\ref{tab:gasfract} gives the fraction of gas in the disc at
given times during the simulation as a function of the initial
corresponding gas fraction value. These
values were obtained as mean values over all the runs which start off
at $t$ = 0 with the given value of the initial gas fraction. The
initial values cover the whole 
range of 0 to 100\%, and becomes 0 to
9\% at $t$ = 5 Gyr and 0 to 7\% at $t$ = 10 Gyr. These  
values are in good agreement with values observed for disc-like
galaxies at intermediate redshifts, as well as with the gas content of
nearby spirals \citep{Erb.SSPRA.06, Leroy.WBBBMT.08, Daddi.MCMWQ.10,
  TacconiP.10, Conselice.MBG.12}.    

\subsection{Early evolution times }
\label{subsec:earlyevol}

\begin{figure} 
\centering
\includegraphics[scale=0.34]{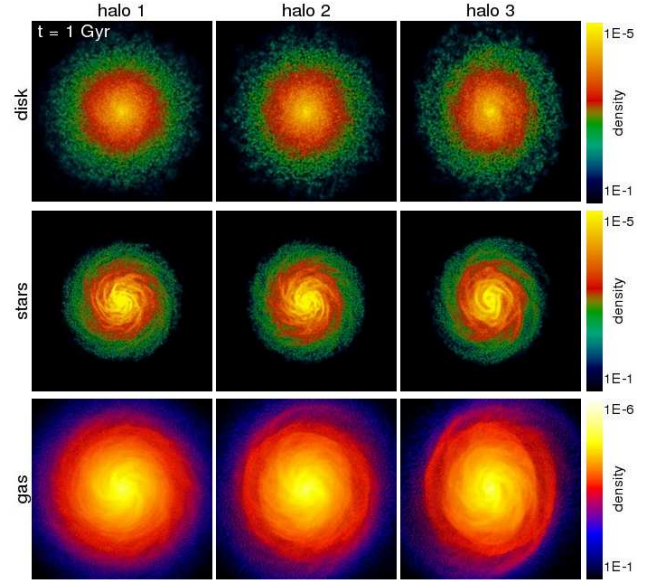}
\caption{Face-on views at $t$ = 1 Gyr for three of our simulations
  with an initial gas fraction of 50\%. 
  The three columns correspond to the three haloes (from left to
  right halo 1, halo 2 and halo 3, and the three rows (from top to
  bottom) to the \textsc{disk}, \textsc{stars} and gas components,
  respectively. Colour coding is according to the local projected
  density of the plotted component,
  as given by the colour bars to the right
  of the plot. The size of each square box corresponds
  to 40 kpc.
}
\label{fig:xy_early}
\end{figure}

Let us first briefly describe the morphology of the disc components
during the early evolution (0 $< t <$ 1 Gyr), i.e. at times when the
bar is not yet visible.

The initial conditions of our simulations were set up so that all
components are as near equilibrium as possible in the common
potential. Therefore, in the cases where the halo is far
from axisymmetric, the \textsc{disk} and gas components are
also initially elongated, but less so than the halo, with initial
$b/a$ values in the relatively outer parts roughly in the range 0.95
-- 0.9 and 0.9 -- 0.8, for  
halo 2 and halo 3, respectively. This lasts at least during the first
Gyr (Fig.~\ref{fig:xy_early}) and in many cases considerably
longer.

Fig.~\ref{fig:xy_early} displays the morphology of the three
components at $t$ = 1 Gyr for the three simulations with 50\% initial
gas. This gives a fair idea of what happens for other initial gas
fractions, since the latter does not influence much this early
evolution, except of course for the amount of stars formed and the
strength of the spiral structure in the stellar component. 
 
The \textsc{disk} component preserves roughly its elongation all
through the early evolution time. Apart from that, it shows no
structure, except, in the case of halo 3, for some faint spiral arms. The gas
component also roughly preserves its elongation during these times. It
develops strong spiral arms of high multiplicity and in a few cases an
inner oval structure. The \textsc{stars} component forms inside out,
i.e. stars form initially at small radii, as expected from the fact that the gas
density is higher in the inner parts. Strong spiral arms can be seen in this
component also, whose positions and sizes correspond well to those of their
gaseous counterparts.  

\subsection{Global trends}
\label{subsec:allviews}

\begin{figure*} 
\centering
\includegraphics[scale=0.96, angle=90]{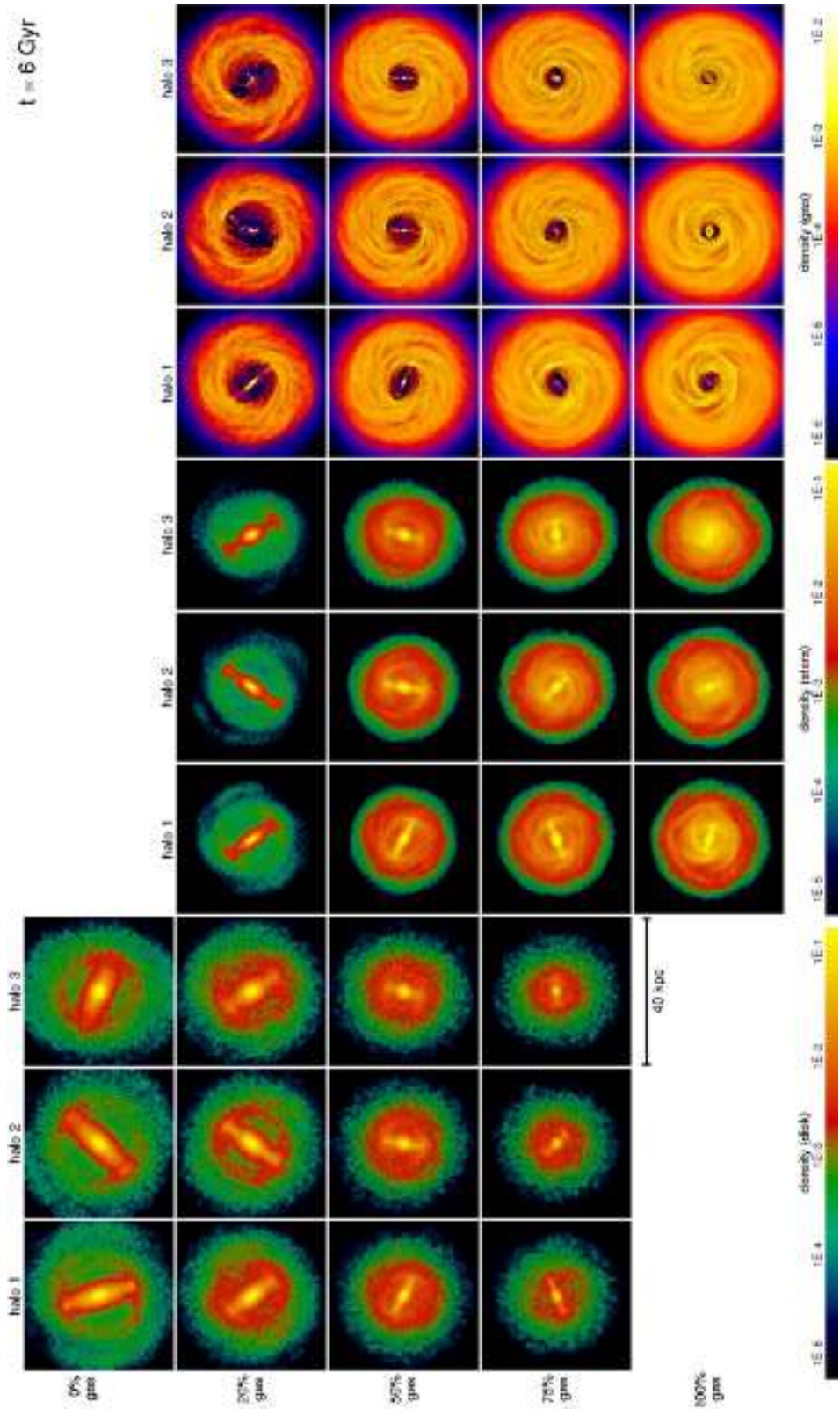}
\caption{Face-on views of the \textsc{disk}, \textsc{stars} and
  gaseous components of our simulations at $t$ = 6
  Gyr. The first three columns correspond to the disc component (old
  stars), the next three to the 
  stars that were formed during the simulation and the last three to
  the gas. The first, fourth and seventh columns show results for
  simulations with halo 1, the second, fifth and eighth with halo 2,
  and the third, sixth and ninth to halo 3. Different rows correspond
  to different initial gas fractions, as indicated in the leftmost
  part of the figure. Rotation is in the
  mathematical sense, i.e. counterclockwise. 
  Colour represents projected density and the
  range is the same for all panels corresponding to the same component
  and the corresponding numerical values are given by the colour bars
  in the bottom of the plot. The size of each square box corresponds
  to 40 Kpc.}
\label{fig:fig_xy_disk_6}
\end{figure*}

\begin{figure*} 
\centering
\includegraphics[scale=0.96,angle=90]{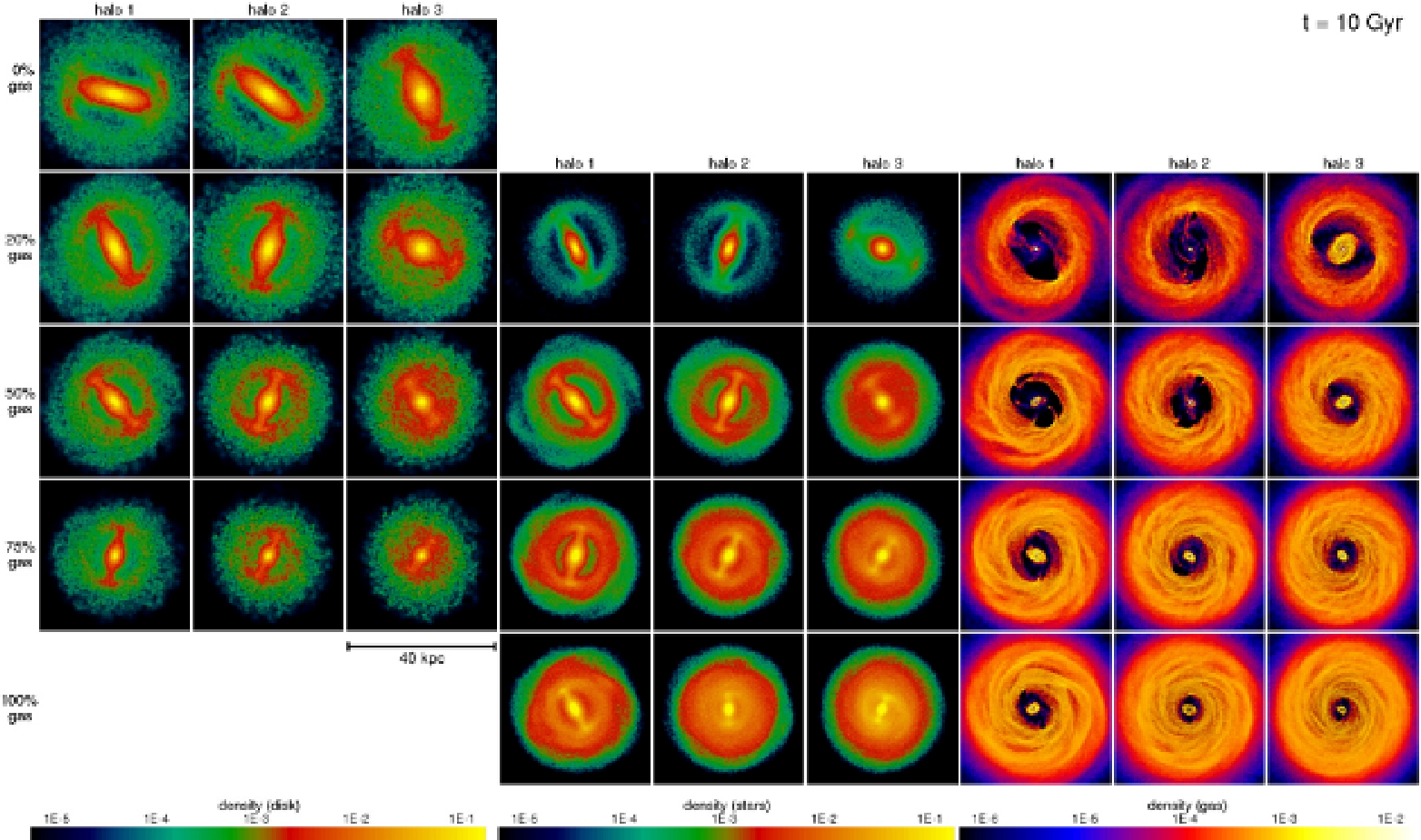}
\caption{As for Fig.~\ref{fig:fig_xy_disk_6}, but at $t$ = 10 Gyr.}
\label{fig:fig_xy_disk_10}
\end{figure*}

Figs.~\ref{fig:fig_xy_disk_6} and \ref{fig:fig_xy_disk_10} display the 
face-on views of the baryonic component of our simulations. The halo
component is not included in the plots and its discussion will be left
for Sect.~\ref{sec:haloprop}. Each figure should be
viewed as three blocks of three columns each. The first block of three
columns corresponds to the \textsc{disk}, the second block
to the \textsc{stars}, and
the last one to the gas. The three columns in each block correspond to 
simulations with the haloes halo 1, halo 2 and halo 3, respectively. Each 
row corresponds to one value of the 
initial gas fraction. From top to bottom these are 0, 20, 50, 75 and
100\%. The simulations initially with no gas will have only the \textsc{disk}
as baryonic component and the simulations with initially a purely
gaseous disc will have no \textsc{disk} component. Thus all blocks of
figures have missing panels. Fig.~\ref{fig:fig_xy_disk_6} gives the
face-on views at time $t$ = 6 Gyr and Fig.~\ref{fig:fig_xy_disk_10}
at $t$ = 10 Gyr. The full evolution of all simulations in all three
components can be seen in our short animations\footnote{
\href{http://195.221.212.246:4780/dynam/movie/gtr}{http://195.221.212.246:4780/dynam/movie/gtr}
}. 

The basic morphological evolution of the \textsc{disk} and \textsc{stars} can be
roughly described as follows: A bar forms, surrounded in the strong
bar cases by a more or less clear ring structure, of
the same extent as the bar, i.e. an inner ring \citep{Buta.95}. 
It should, however, be kept in mind that these figures do not cover all
possible morphologies, since the basic halo and total disc
profiles are the same in all cases, and all that changes is the halo
shape and the gas fraction.   

The gas morphology is reminiscent of what was found in the earlier
simulations of \cite{Athanassoula.92b}, i.e. considerably different from
that of the \textsc{disk} and \textsc{stars}; the most striking
difference being the absence of a bar in the gas component. In the
centre there is a strong concentration of gas, which we will hereafter
call the gaseous central mas concentration (CMC). This is surrounded by a
large very low density annulus, whose inner and outer radii are a
function of the simulation parameters and the time. The extent of
this annulus increases with time, as can be seen by comparing
Figs.~\ref{fig:fig_xy_disk_6} and
\ref{fig:fig_xy_disk_10}. Surrounding this region, there is 
the disc of gas, in which there are clear spiral segments, but no
clear-cut two armed global spirals. 

Looking carefully, one can discern in the
very low density annulus two thin stripes of gas, linking the CMC to
the gas disc surrounding the very low density annulus. Their location
with respect to the bar, as well as their extent, links them to the gas
concentrations in the shocks on the leading edges of the bar, found
e.g. in the purely hydrodynamical simulations of
\cite{Athanassoula.92b}. They are of course much less symmetric and less
well outlined than in those simulations, but this should be expected
since the older simulations were response simulations of an
isothermal gas in a rigid model bar galaxy and did not include
self-consistency, the physics of the gas, star formation, or feedback.
It is, nevertheless, clear that it is the same features in the two cases. 
We examined these features also in our high resolution simulations,
with up to 20 times more gas particles and up to five times more
\textsc{disk} and halo particles. We find that the morphology is similar, 
although one can discern more details of the flow, and the gaseous
density concentrations along the leading edges of the bar are much
clearer delineated.

Turning now to the \textsc{disk} and \textsc{stars} components and
comparing times $t$ = 6 and 10, we see that in the latter time the bar is
longer and stronger, as expected due to secular evolution
(e.g. A03), and the inner ring is more clearly defined. 

Both at $t$ = 6 and at $t$ = 10 Gyr, there is a clear gradient in bar
strength from top left to bottom right in the left and middle block of
panels. The 
strongest bars are found for the spherical halo and no gas, and the
strength decreases as we go towards initially more triaxial haloes and larger
gas contents. This will be established quantitatively in
Sect.~\ref{sec:barstrength}, but can already be qualitatively seen in
Figs.~\ref{fig:fig_xy_disk_6} and \ref{fig:fig_xy_disk_10}. Also the 
extent of the low density region in the gas component follows a
related trend along the same 
diagonal. Namely, it is largest for spherical haloes and minimum gas content
and decreases as the initial halo triaxiality increases and/or as the
initial gas fraction increases. 

Another such gradient is linked to the outer extent of the projected
surface density in all components. This, however, does not necessarily
reflect the concentration of each component, but more its total mass.
Indeed, we use the same colour coding for all panels of a given
component (and the same for the \textsc{disk} and the \textsc{stars}. Thus,
the surface density of the \textsc{disk} will be higher for runs with
initially less relative gas, while the surface density of
the \textsc{stars} will be higher for runs with an initially higher relative gas
content. This could explain the trends seen in
Figs.~\ref{fig:fig_xy_disk_6} and \ref{fig:fig_xy_disk_10}. Namely 
the old disc seems
most extended for the simulations with no gas and its extent decreases
as the initial gas fraction increases.  The opposite is true for the
gas and the stars formed during the simulation. 

Presumably due to the gaseous component and the corresponding CMC mass, we
obtained in our simulations morphological features which 
had not been seen so far in pure N-body simulations. In particular,
we note an oval (in some cases near-circular) component of high
density in the central part of most of the simulations seen
face-on. It is quite clear in both the \textsc{disk} and the
\textsc{stars} components of simulations with e.g. initially 50\% gas
and its size is of the order of one third of that of the bar
(Figs.~\ref{fig:fig_xy_disk_6} and \ref{fig:fig_xy_disk_10}). At
first sight one could mistake this for a classical bulge. For our
simulations, however, it is clear that this can not be true, since, by
construction, they have no classical bulge. An alternative would be a
pseudo-bulge \citep[as named by ][]{Kormendy.Kennicutt.04} or discy bulge
\citep[as named by ][]{Athanassoula.05b}. A third alternative 
interpretation would be to link this component to the barlens, which
was discussed recently by \cite{Laurikainen.SBK.11} when they studied
the central region of many NIRS0S galaxies. This is denoted by {\it
  bl} and is generally distinguished from nuclear lenses by its much
larger size. Typical examples of a {\it bl} can be seen in NGC
2983 (figure 8 of \citealt{Laurikainen.SBK.11}) and in NGC 4314 (figure 9
of the same paper, where the fine-structure in the central regions
confirms that this component cannot be a bulge). The nature of these
components, their formation mechanism, as well as their properties will
be discussed elsewhere.  

\subsection{Morphology of the gaseous CMC at the late simulation times }
\label{subsec:CMClate}

\begin{figure}
\includegraphics[scale=0.14]{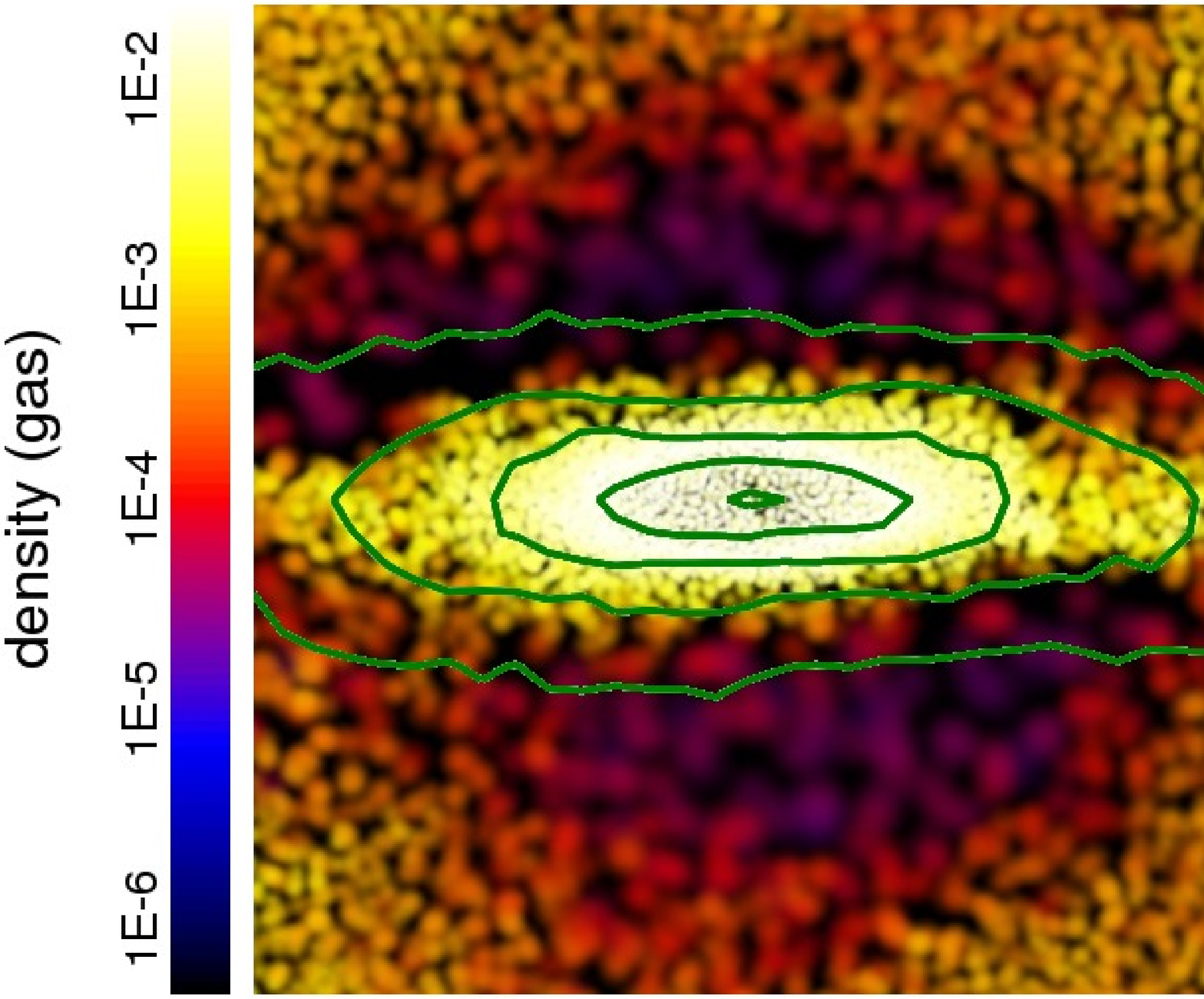}
\caption{
Gas density in the central region of two simulations. The left panel
shows simulation 109 
at time 2.7 Gyr, the middle one the same simulation at time 5.07 Gyr
and the 
right panel simulation 115 at time 9.04 Gyr. Colour is according to the
projected density as given by the colour bar on the left of the
panels. The snapshots are 
rotated so that the bar is horizontal and the size of the boxes
is 8 $\times$ 8 kpc. The green lines on all three panels show the
isophotes for the combined \textsc{disk} and \textsc{stars} components.} 
\label{fig:gasCMC}
\end{figure}

Fig.~\ref{fig:gasCMC} gives a detailed view of the central region of
two simulations and shows clearly that the gaseous CMC has a
rather complex structure. It has a small, high density inner
part, which can be clearly seen as a central white feature in this
figure. This is 
elongated roughly along the bar. It forms early on in the evolution and
is usually larger at early times than at later ones. Its size and
axial ratio
vary considerably with time. Eye estimates give a representative
outline of 0.75 by 0.6 kpc, but at late times it can be considerably
smaller, while at early times it can be as large as 
1.8 by 0.8 kpc. It is particularly clear in simulations with
initially 20\%, or 50\% gas where it is clearly seen to form first, before the
outer component. In the initially 20\% cases this component has a
rather interesting 
morphological evolution. It is already present at $t$ = 2.5 $\pm$ 0.5 Gyr,
where it can be seen as a rather extended component. Its central density
is considerably lower than that of its outer part and it can therefore be
considered as an elongated inner ring. As it evolves, it becomes
smaller and rounder and has almost acquired its final shape and extent
by  $t$ = 5 $\pm$ 1 Gyr. Over some part of the simulation this could
therefore be considered as either an inner ring, or an inner gaseous
bar, as has been found in previous simulations \citep{Heller.SE.01}. 

The gaseous CMC has also a second, considerably more extended component (yellow
in Fig.~\ref{fig:gasCMC}). This has lower density and is oriented roughly
perpendicular to the bar. Its outline is less well defined than that
of the (white) inner component and is more irregular. Its typical size
is between 1 and 3 kpc, but in some cases can be even
larger. It is not very elongated, with, in many cases, an axial ratio
of the order of 2:3. It forms considerably later than the
white inner component. In general, it forms earlier for runs with
initially strongly triaxial haloes (halo 3) and later in simulations
with an initially more spherical halo. There is also a general trend between
the initial gas fraction and the time at which this component forms,
in the sense that it forms earlier in more gas rich cases. In fact it
has not formed by the end of the simulation ($t$ = 10 Gyr) for runs
106 and 109 which have initially only 20\% 
gas and spherical or mildly triaxial haloes, respectively, and it
forms only after 9 Gyr for runs 111 and 114 which have initially 50\%
gas and the same haloes.

\section{Bar strength evolution}
\label{sec:barstrength}

\begin{figure}
\includegraphics[scale=0.5]{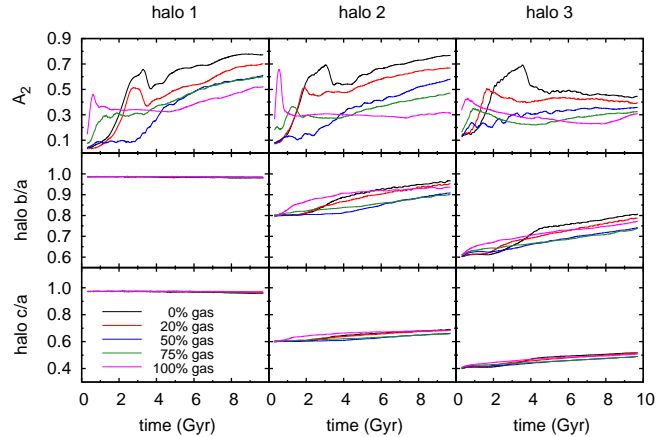}
\caption{Time evolution of three quantities. From top to bottom the
  rows of plots give the bar strength ($A_2$), the halo axial ratio in
  the equatorial plane ($b/a$) and the halo flattening ($c/a$). The
  three columns correspond to the three different halo models. In the
  online version, the different
  initial gas fractions are shown with lines of different 
  colour, namely black for 0\%, red for 20\%, blue for 50\%, green for
  75\% and magenta for 100\% gas, as explained in the panel of the
  first column and third
  row. In this display one can easily see the effect of the initial 
  gas fraction on the results. 
  }
\label{fig:A2-boa-coa}
\end{figure}

\begin{figure*}
\includegraphics[scale=0.74, angle=-90]{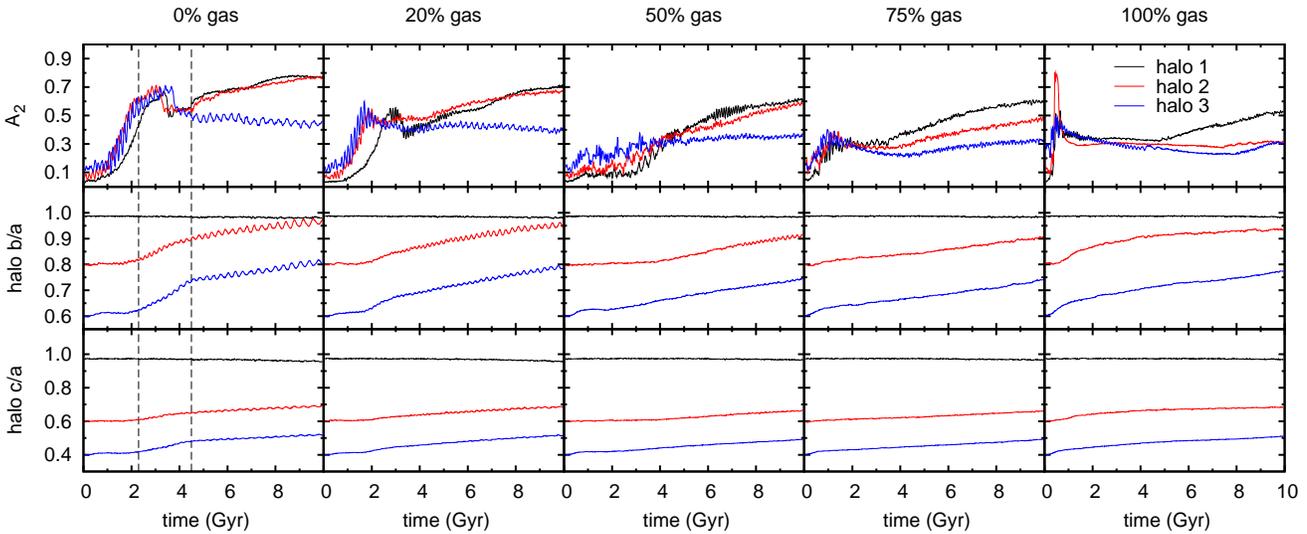}
\caption{This figure shows the same information as
  Fig.~\ref{fig:A2-boa-coa}, but now the data is displayed so as to show best
  the effect of halo triaxiality. From top to bottom the rows of plots
  give bar strength, the halo axial ratio in the equatorial plane and the
  halo vertical flattening. The five columns
  correspond to the five different initial gas fractions and the
  different haloes are shown with lines of different styles and
  colour, as given in the upper right panel. The black dashed vertical
  lines in the first column of panels outline  
  the time range during which the $b/a$ axial ratio displays a strong
  growth.
} 
\label{fig:A2-boa-coa_alt}
\end{figure*}

The time evolution of the bar strength is shown for all runs in the
upper panels of Fig.~\ref{fig:A2-boa-coa}. It was obtained as
described in Sect.~\ref{subsec:analysis} and then smoothed with a
Savitzky-Golay filter \citep{Press.TVF.92}. We wilfully chose filter
values that smooth out not only the 
noise, but also some relatively high frequency oscillations which we
will discuss in detail in Sect.~\ref{sec:oscillations}, so as to view only
the global evolution. The 
evolution for the various simulations is displayed here so as to show
best the effect of the initial gas fraction. 
Fig.~\ref{fig:A2-boa-coa_alt} displays the same data as
Fig.~\ref{fig:A2-boa-coa}, but now so as to reveal best 
the effect of halo initial triaxiality. Moreover, in
Fig.~\ref{fig:A2-boa-coa_alt} we used much less smoothing,
smoothing out only what we verified by eye is indeed noise. From these
two figures it becomes clear that the effect of the initial gas
fraction is very crucial, even more so that the effect of the initial halo
triaxiality. Note also that their effect prevails at different
times. More specifically we can say the following: 

The time at which the bar starts forming depends considerably on the
halo triaxiality (Fig.~\ref{fig:A2-boa-coa_alt}) and must be
presumably due to the triggering by the halo
non-axisymmetry. This effect of triaxiality can be clearly
seen in all cases, except those with a strong initial gas fraction
where initial disc instabilities in the inner disc parts and the
formation of an inner bar do not allow us to distinguish when the
main bar starts growing. 

All models with initially no gas and the model with 20\% gas and a spherical
halo (run 106) have the same {\it four}
evolutionary phases, independent of their halo shape: a fast growth
phase, followed by a plateau-like part and then a sharp decrease (see
also MA10). The fourth and final phase is that of a slow secular evolution. The
maximum values after the growth phase are roughly the same in all
gas-less runs and so is the amount of decrease after the 
plateau, while the times at which these features occur changes little 
between runs. 

The remaining models have fewer evolutionary phases and in many
cases it is difficult to distinguish between them.
The two simulations with initially 20\% gas and a triaxial halo have a similar
time evolution of the bar strength, which resembles the one described above,
but lacks the plateau right after the phase of the bar growth, while
the drop is not as clear-cut. We can thus say that there are three
evolutionary phases, first a  
bar growth to a maximum value, followed by a short decrease
phase and finally a slow secular evolution phase. 
 
For all simulations with a strong initial gas fraction (50\% or higher)
the $m$ = 2 strength curves are simpler, and have fewer
evolutionary phases. In particular for the cases with initially 50\% gas and a
spherical, or mildly triaxial halo there is first a time interval 
during which axisymmetry prevails, followed by a time of bar
growth. Both these time intervals are much longer than in the
gas-less or gas-poor cases described above, so that we can describe
this growth as {\it secular growth}. This is followed, as in
the previously described runs, by a secular evolution phase.
 
\begin{figure}
\includegraphics[scale=0.4,angle=-90.]{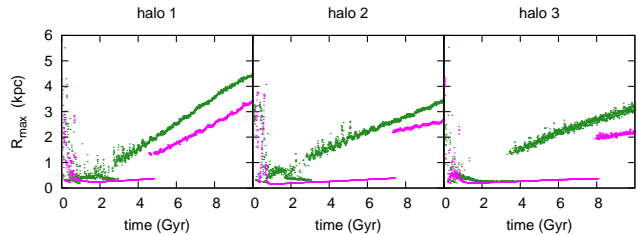}
\caption{$R_{max}$, the radius at which the relative $m$ = 2 amplitude
  is maximum, as a function of time. The green points are for a
  simulation with initially 75\% gas and the magenta ones for one with
  initially 100\% gas. From left to right the three panels correspond
  to halo 1, halo 2 and halo 3, respectively.}
\label{fig:rmax}
\end{figure}

Understanding the evolution of gas-rich simulations (initially 75\%, or 
100\% gas) is more complex and we need information on the time
evolution of $R_{max}$, the radius at which the $m$ = 2 relative
Fourier component is maximum,
which we give in Fig.~\ref{fig:rmax}. Viewing animations of the
evolution of the disc component we see that in the first Gyr or so,
several noisy, short-lived, non-axisymmetric features develop in the
\textsc{stars} component which, since these stars are recently formed, 
is still quite cold (i.e. it has a low velocity dispersion). These 
features can be spirals, distorted bars, or rings and contribute a lot
of noise to the results in Fig.\ref{fig:A2-boa-coa} to \ref{fig:rmax} in the
very early times. A particularly strong such feature contributes a
very strong and narrow peak of the $m$ = 2 strength for run 120 (see
Figs.~\ref{fig:A2-boa-coa} and ~\ref{fig:A2-boa-coa_alt}).

As these features subside, a very short bar forms, which we can call
an inner bar. The corresponding values of $R_{max}$  are very small
(Fig.~\ref{fig:rmax}). The main bar 
starts growing well after the inner bar and, when it becomes sufficiently 
strong it provokes an abrupt jump of the value of $R_{max}$ to a
considerably larger value, compatible with what one would expect for a
main bar. Fig.~\ref{fig:rmax} shows clearly that this jump occurs at
later times for runs with higher gas fractions and more prolate
haloes, in good agreement with the results we presented already for
simulations with initially up to 50\% gas and with the more
qualitative impression one gets from the face-on evolution
animations. Thus, in gas-rich cases one has to consider the $A_2(t)$
curves as a result of the growth and evolution of two components combined, one
which can be called an inner bar and occurs earlier, before the bar
itself has grown, and a second one which is the bar.     

All evolutions, for all gas fractions and for all halo types, end with
a phase of slow secular evolution, whose duration and strength varies
from one simulation to another. In this phase the effect of
the halo shape is very strong. In spherical halo cases, and for all
initial gas contents, there is strong secular evolution, as
witnessed by the slope of the corresponding $A_2(t)$ curves. At the
other extreme, halo 3 cases (initially strongly triaxial) have at the
best a mild secular evolution (Fig.~\ref{fig:A2-boa-coa_alt}). Halo 2
cases are intermediate, more similar to spherical cases in gas-less or
gas-poor cases and more similar to the strongly triaxial cases in 
gas-rich ones. 

There is thus a duality in the effect of the
non-axisymmetric forcing of the triaxial halo on bar evolution. In the
very early times this forcing can trigger 
bar formation, so that bars in triaxial haloes grow earlier than in
spherical ones, as already discussed in the beginning of this
section. On the contrary, at the later stages of evolution, when 
the bar is well grown, triaxiality hinders bar growth due to
the nonlinear interaction between the two non-axisymmetries. Indeed,
such an interaction could induce chaos, as advocated by
\cite{ElZant.Shlosman.02}. Testing this, however, is not 
straightforward since one should be careful about eliminating the
contribution of `confined chaos' from the statistics, because this can
account for galactic structures even for timescales of the order of a
few Gyr (see discussion in \citealt{Athanassoula.RGBM.10}).  

\section{Are bars long-lived or short-lived?}
\label{sec:longshortlived}

\subsection{Context}
\label{subsec:longshortlived-context}

Are bars in isolated galaxies long-lived, or short-lived? A massive
central black hole, or a CMC can destroy a bar, provided it is
sufficiently massive and/or centrally concentrated \citep{Hasan.PN.93,
  Norman.SH.96}. More recent work, however, has shown that the
required values are too high compared to those of observed CMCs \citep*{Shen.Sellwood.04,
  Athanassoula.LD.05}.  

The debate became more animated when \cite{Bournaud.Combes.02}
reported a number of simulations which included a gas component and in which
the bar was destroyed and then formed anew with
the help of gas accretion, only to be destroyed again. Three or
four such bar episodes occurred during each run. 
\cite{Shen.Sellwood.04} criticised these simulations for having a time
step too long to properly describe the orbits in the
central region, so that the bar destruction would be due to an
inadequacy of the simulations rather than to a physical effect. In
response to this, (\citealt{Bournaud.CS.05}, hereafter BCS) decreased the
time step of their simulation by a
factor of 8 and still found that the bar was destroyed. Using
specifically designed simulations, they argue that the effect of the
CMC only is indeed not sufficient to fully dissolve the bar. On the
other hand, the role of the gas in the angular momentum exchange
within the galaxy has a much stronger effect on the bar strength and
can indeed destroy it, particularly when it is added to the effect of
the CMC. They thus conclude that bars are transient features with a
lifetime of 1-2 Gyr.
                        
\cite{Debattista.CMM.04,Debattista.MCMWQ.06} contributed to this debate
by running a number of simulations with either rigid or live haloes
and no gas, and, in all but one cases, 
they witnessed that the buckling instability weakened the bar, but did
not destroy it. The exception included
gas cooling, but no star formation and no feedback thus resulting
in the formation of a particularly massive and compact CMC which
destroyed the bar. 
The \citeauthor{Debattista.MCMWQ.06} result
thus disagrees with 
the BCS one, because it is the CMC that drives the bar destruction for
the simulation of the former and the angular momentum exchange for
that of the latter. 

\cite{Berentzen.SMVH.07} also used simulations to
examine this issue -- including specifically designed ones, like those
in BCS -- but did not find bar destruction. They argue that what BCS 
witness is simply the decrease of the bar amplitude due to its
buckling. In the \cite{Berentzen.SMVH.07}
simulations (as well as in the later simulations of
\protect\citealt{Villa.VSH.10}) the amplitude of the bar is indeed decreased due
to the buckling, but the bar is not destroyed, and after the buckling the
bar amplitude starts increasing again.

\subsection{Input from our simulations}
\label{subsec:longshortlived-ours}

There are many technical differences between the
simulations of the two groups. \cite{Bournaud.CS.05} have a large
number of gas particles ($10^6$), model the gas with sticky
particles and use a rigid halo. The latter may have particularly
important consequences, because,
as shown by \cite{Athanassoula.02}, a rigid halo can not take
part in the angular momentum redistribution and thus can not help the
bar grow. On the hand, \cite{Berentzen.SMVH.07, Villa.VSH.10} have a
live halo and an SPH gas description, but the number of gas particles
in their simulations is 
rather low, 40\,000 only, and they do not undergo any star
formation, feedback or cooling. 

The simulations we describe in this paper can be used to shed new
light on this important and not yet settled issue.
Our simulations have a large number of gas particles
(Sect.~\ref{subsec:simu}) and we have also looked at the bar
strength evolution in the three simulations with a yet higher number of
particles (up to 4 million gas particles). Furthermore, all our
simulations have a live halo 
with a sufficient number of particles to describe the resonances
adequately \citep{Athanassoula.02} and their softening is 50 pc,
thus ensuring a high spatial resolution. All previous simulations
concerning bar longevity were performed with spherical haloes, so for
comparisons we will restrict ourselves to our simulations with
spherical haloes. Both the animations of these
simulations and the plots of the bar strength time evolution 
(Figs.~\ref{fig:A2-boa-coa} and \ref{fig:A2-boa-coa_alt}) show
clearly 
that for the spherical halo cases the bar is never destroyed, and this
for all gas fractions. The drop of the $m$ = 2 amplitude
observed at very early times in simulations whose disc is initially
all gas, or very gas rich (Fig.~\ref{fig:A2-boa-coa} and
\ref{fig:A2-boa-coa_alt}) is due to the demise of initial gas
instabilities which are reflected in the \textsc{stars} component.
As already discussed in Sect.~\ref{sec:barstrength}, these occur in
the inner parts of the disc. Thus one can not exclude the dissolution
of inner bars with a length of the order of, or considerably less than
1 kpc. 

We thus conclude that our simulations argue against bar destruction 
and agree with those of \cite{Debattista.CMM.04,Debattista.MCMWQ.06,
  Berentzen.SMVH.07} and \cite{Villa.VSH.10}, even though the halo
radial profiles and the way the gas is modelled varied strongly from
one set of simulations to another. 

We can, furthermore extend this discussion to triaxial haloes.
Fig.~\ref{fig:A2-boa-coa} shows that, even in simulations with a
triaxial halo, the bar does not get destroyed.
although its $m$ = 2 amplitude in the case with initially 100\% gas 
starts and stays small. In some of the simulations with an initially strongly
triaxial case (halo 3), however, the $m$ = 2 amplitude shows a very small
decrease with time. As can be seen by comparing
Figs.~\ref{fig:fig_xy_disk_6} and \ref{fig:fig_xy_disk_10} this is not
due to a weakening of the bar, but to the growth of the barlens
component (Sect.~\ref{subsec:allviews}). Furthermore,
this weakening is slow that it would have a sizeable effect only at
times much longer than the a Hubble time.

\section{Halo properties}
\label{sec:haloprop}

\subsection{Radial profiles of halo equatorial axial ratios}
\label{subsec:halo-bovera}

\begin{figure}
\includegraphics[scale=0.45]{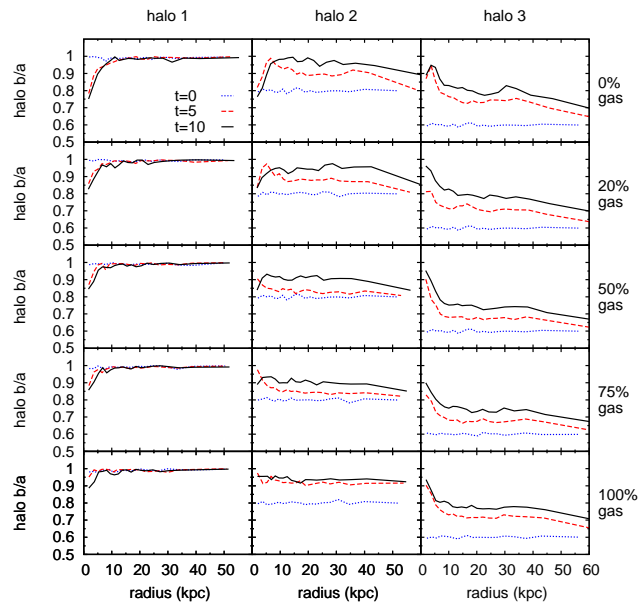}
\caption{Radial profile of the halo equatorial axial ratio ($b/a$) for
  three times during the simulation ($t$ = 0, 5 and 10 Gyr) and for
  all simulations. The layout is as for Fig.~\ref{fig:vcirc}}.
\label{fig:shaperadius}
\end{figure}

Fig.~\ref{fig:shaperadius} shows the radial profile of the halo
equatorial axial ratio ($b/a$) for three times ($t$ = 0, 5 and 10
Gyr) and for all simulations. In simulations with an initially
spherical halo (halo 1) and no gas, 
the innermost part becomes during the evolution considerably
triaxial, thus forming a bar in the halo component. This structure was
already spotted in a number of simulations
\citep[e.g.][]{Debattista.Sellwood.00, ONeil.Dubinski.03,
  HolleyBockelmann.WK.05, Berentzen.Shlosman.06} and its properties have
been studied in some detail in \cite{Hernquist.Weinberg.92},   
\cite{Athanassoula.05b, Athanassoula.07} and \cite{Colin.VK.06}. It is
called `halo bar' or `dark bar' and is considerably 
shorter than the disc bar, rotates with roughly the same angular
velocity and is due to the angular momentum exchange between the
near-resonant particles in the inner halo and the near-resonant particles 
in the disc bar region \citep[A03]{Athanassoula.07}. 

This feature is also seen in all our simulations with a spherical halo
and its amplitude depends on the gas fraction, being the strongest in
gas-less simulations. It is also seen in cases with a mildly triaxial 
halo (halo 2) and with an initial gas fraction up to $\sim$50\% and, albeit
it to a much lower extent, in the simulation with the strongly triaxial
halo (halo 3) and no initial gas. The other 
simulations do not have this feature and, on the contrary, show an
increase of $b/a$ at small radii. The initial gas fraction that limits the
simulations with a halo bar from those with a rounder centre depends
on the halo shape and is in fact between 50 and 75\% for the initially
less triaxial halo and around 0\% for the strongly triaxial one.  
It is clear that the existence of this feature is linked with the
existence of a strong bar in the disc component, since the snapshots
that do not have it either have no bar, or a only a very weak one.   
We find a rough limit of $A_2 > 0.5$ for the halo bar to exist.

There is also a rather abrupt drop of $b/a$ in the outermost parts of
the halo. Although the halo is indeed less spherical in its
outermost parts, the abruptness of the transition is an
artefact due to the way the $b/a$ values were obtained. Indeed as
discussed in Section~\ref{subsec:analysis}, in order to calculate the 
inertial tensor the particles were binned by their local density value
in bins of equal number of particles and, since the density drops
considerably with radius, the radial extent of the outer bin is much larger
than that of the others.   

Barring this drop of $b/a$ in the innermost and the outermost parts,
globally the halo evolves towards axisymmetry.
This time evolution can be better seen in the second row 
of panels in Figs.~\ref{fig:A2-boa-coa} and \ref{fig:A2-boa-coa_alt}
and is discussed in Sect.~\ref{subsec:evolfaceon}.    

\subsection{Time evolution of the halo axial ratios}
\label{subsec:evolfaceon}

Fig.~\ref{fig:A2-boa-coa} shows the time evolution of $b/a$, the halo 
axial ratio in the equatorial plane, (middle panels) and $c/a$, the halo
flattening (lower panels). The data are displayed so that they show best
the effect of initial gas fraction. 

Changes of the axial ratios in the spherical haloes are very small, 
of the order of a few per cent over 10 Gyr. They show very little dependence
on the gas fraction. This small gradual flattening,
i.e. small decrease of the $c/a$ ratio with time, is not due to the
simulation starting somewhat off equilibrium. This would have led to a
fast rearrangement and not a gradual, slow evolution. Also many tests
have shown us that the iterative method we use \citep{Rodionov.AS.09,
  Rodionov.Athanassoula.11} can produce initial conditions very near
equilibrium for times sufficiently long for us to be able to follow bar
formation uninhibited by other instabilities due to the inadequacy of
the initial conditions. Nevertheless, the evolution of the galaxy due
to the formation and evolution of the bar will influence the halo and
can well account for the small changes we see in its axial ratios.

On the other hand, simulations with an initially triaxial halo, have a
much stronger time evolution of their $b/a$ and $c/a$
values than simulations with initially spherical halo, as expected. 
In simulations with no gas and with an initially triaxial halo we
confirm that the halo becomes considerably less triaxial during 
the evolution, as already found by 
\cite{Berentzen.Shlosman.06}, \cite{Heller.SA.07} and MA10.
Our results allow us to extend this conclusion to simulations with
gas. The corresponding haloes also become more spherical during the
evolution, but the effect is smaller than for the gas-less
models. Nevertheless, a systematic dependence on the gas fractions
is inconclusive, as for example in the middle right panel of
Fig.~\ref{fig:A2-boa-coa} (halo 3) the initially 100\% and 20\% gas end up with
roughly the same $b/a$ ratio, which is larger than that of the 50\% and 70\%
gas cases. The changes in $b/a$ are accompanied by a change in 
$c/a$, which is nevertheless much smaller than the $b/a$ change. Both
together bring the haloes nearer to sphericity. Also the amount of gas
has less effect on the change of the $c/a$ than on that of the $b/a$. 

The middle and lower panels of Fig.~\ref{fig:A2-boa-coa_alt} display
the same data, but so as to show best the effect of the halo. The
$b/a$ for halo 2 starts from 0.8 and ends roughly in the range 0.9 to 0.98,
while for halo 3 it starts from 0.6 and ends roughly in the range 0.75
to 0.8. The corresponding numbers for $c/a$ are from 0.6 to 0.65/0.7 for
halo 2, and from 0.4 to 0.5 for halo~3.

This way of plotting the data also allows us also to see best the imprints
of the various phases of bar formation and evolution on the temporal
evolution of the $b/a$ and $c/a$ profiles. For simulations where the
bar growth has four phases (see Sect.~\ref{sec:barstrength}), the
$b/a$ growth has clearly three phases. A first phase where the $b/a$
value hardly changes, a second phase with a strong growth and finally a
third phase with a weaker secular growth. To make this yet clearer,
we added on Fig.~\ref{fig:A2-boa-coa_alt} vertical lines 
roughly delineating these three phases for the 0\% gas case. By
extending them to the upper panel ($A_2$ as a function of time) we see
that the first phase corresponds to the bar growth phase, the third one
to the bar secular evolution phase, and the intermediate phase
encompasses the flat $A_2$ phase, the abrupt fall of $A_2$ and the
very first steps of the bar secular evolution. The three phases in the
$b/a$ evolution are clearly seen only in those runs where the $A_2$
has four phases (see Sect.~\ref{sec:barstrength}). Short duration
strong growths are also seen in other cases and are again linked 
to specific $m=2$ phases. For example in initially 100\% gas cases
there is a strong $b/a$ growth roughly in the first Gyr, i.e. the time
when the strong $m=2$ features appear. Thus the $b/a$ growth is
clearly linked to the bar formation and evolution, as already proposed
in MA10. 

\subsection{Angular momentum redistribution}
\label{subsec:angmom}

\begin{figure}
\centering
\includegraphics[scale=0.35, angle=-90.]{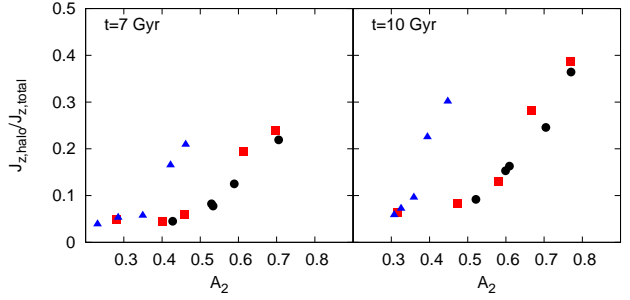}
\caption{Correlation between the fraction of angular angular momentum
  absorbed by the halo and the bar strength. Symbols 
represent different halo shapes: filled circles (black online) for
halo~1, filled squares (red online) for halo~2 and filled triangles
(blue online) for halo~3. 
}
\label{fig:A2Jhalo}
\end{figure}

Previous N-body simulations have shown that barred galaxies evolve by
redistributing their angular momentum and that haloes 
play a substantial role in this redistribution by absorbing angular
momentum at their various resonances, mainly the inner Lindblad
resonance, the corotation resonance and the outer Lindblad resonance
\citep{Athanassoula.02}. Furthermore the fraction of the initial
angular momentum that is absorbed by the halo correlates well with the
bar strength (A03).  

Here we extend this to simulations with gas and/or with triaxial
haloes. The results are given in Fig.~\ref{fig:A2Jhalo} for $t$ = 7
(left panel) and 10 Gyr (right panel). We note that simulations with
an initially 
spherical halo (halo 1) share the same trend as those with an
initially mildly triaxial halo (halo 2), while simulations with an
initially strongly triaxial halo (halo 3) have considerably higher
halo angular momentum exchange, at least for the two cases with the
strongest bars. This must be linked to the extra torque due to 
the triaxial halo, but its study is beyond the scope of this paper.
The Pearson correlation coefficient \citep{Press.TVF.92} for the
simulations with halo 1 
or halo 2, taken together, is 0.89 (0.90) and for those with halo 3
it is 0.93 (0.97) for time 7 (10) Gyr. 

\subsection{Halo kinematics}
\label{subsec:halokinematics}

\begin{figure}
\centering
\includegraphics[scale=0.35, angle=-90.]{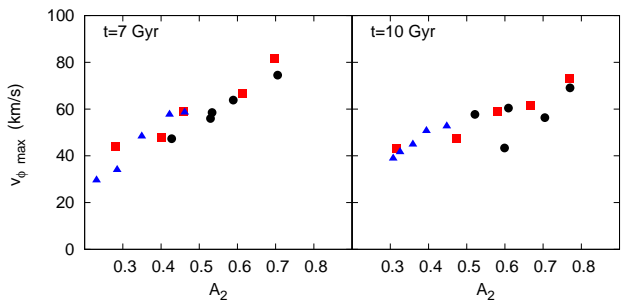}
\caption{Correlation between the peak tangential velocities of
the halo permanently disc-like particles (see text) and bar strengths. Symbols 
represent different halo shapes: filled circles (black online) for
halo~1, filled squares (red online) for halo~2 and filled triangles
(blue online) for halo~3. 
}
\label{fig:halovphi_A2}
\end{figure}

The angular momentum redistribution within the galaxy brings
significant changes to the halo velocity 
distributions \citep{Athanassoula.05a, Athanassoula.07, Colin.VK.06}.
Halo particles located relatively
near the disc equatorial plane acquire tangential velocity and rotate
in the same sense as disc particles,
albeit much slower \citep{Athanassoula.07}. This result was extended to
initially non-spherical haloes by MA10. We will here examine how
the presence of gas can influence these results. 

We follow \cite{Athanassoula.07} and select
all particles that remain near the equatorial plane for a considerable time
interval and call them `disc-like' for brevity. In practice, we choose particles
for which $|z|<2$~kpc for 
all times within the range $7<t<10$~Gyr (for $t$ = 10 Gyr) and
$4<t<7$~Gyr (for  $t$ = 7 Gyr). The resulting velocity radial profiles
of these particles are very similar to those shown in Fig. 11
of \cite{Athanassoula.07}, or Fig. 23 of MA10 and we thus do
not display them here. They show that mean tangential 
velocity curves for such `disc-like' halo particles can reach
velocities of the order of 80 km/sec. They also show 
that this rotation is fastest for the
spherical model with no gas, which has the strongest bar, and slowest
for the model with initially the most gas and most triaxial
halo, i.e. follows a trend similar to that of the bar strength. 

In order to illustrate this trend, we plot in
Fig.~\ref{fig:halovphi_A2} the peak tangential velocities of the
`disc-like' halo particles versus the disc bar strengths of the 
corresponding model for $t$ = 7 and $t$ = 10~Gyr. It is clear that
models with stronger bars have higher peak rotation. In fact, the
corresponding correlations are quite strong, with Pearson correlation
coefficients \citep{Press.TVF.92} of
0.95 and 0.86, for $t$ = 7 and 10 Gyr, respectively. This clearly
links the origin of these 
tangential velocities to the bar and is in good agreement with
results of \cite{Athanassoula.07} and MA10. 

In Fig.~\ref{fig:halovphi_A2}, contrary to the results in
Fig.~\ref{fig:A2Jhalo}, 
the simulations with an initially strongly non-axisymmetric halo (halo
3) lie on the same correlation as those of 
halo 1 and halo 2. This argues that there is not a one to one
correspondence between the halo population that absorbs the angular
momentum and the population with the largest $v_{\phi}$. Indeed in the
latter there are many particles that already initially
were considerably rotating. An in depth analysis of the orbital
structure in the halo and its evolution as angular momentum is
absorbed will be given elsewhere.

\subsection{Halo bulk rotation}
\label{subsec:halobulkrot}

\begin{figure}
\centering
\includegraphics[scale=0.75]{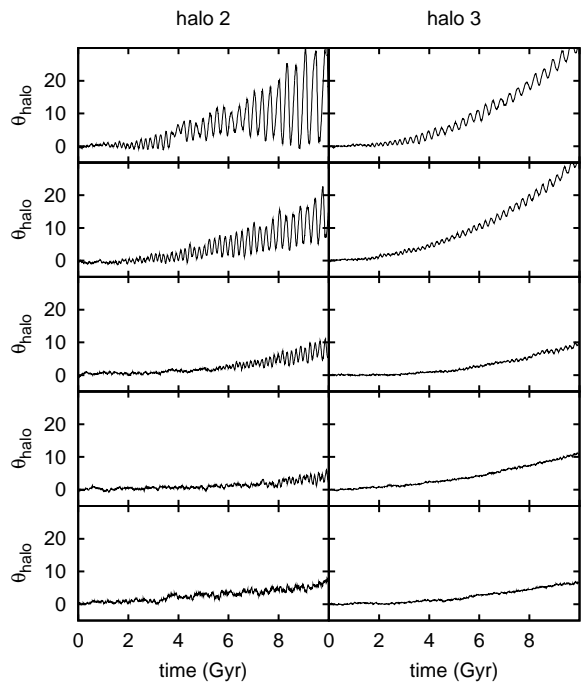}
\caption{The angle of the halo major axis as a function of time for the 10
  simulations with non-axisymmetric haloes. The left (right) column displays
  information on simulations with initially mildly (strongly) triaxial
  haloes. From top to bottom the initial gas fraction is 0\%, 20\%, 50\%, 75\%
  and 100\%. 
}
\label{fig:halovphi_angle}
\end{figure}

Triaxial haloes are not simple geometrical objects. As discussed in
Sect.~\ref{subsec:evolfaceon}, the inner part
presents an elongation, which we called halo bar and which is rotating
together with the bar in the disc. Further out also the halo is
non-axisymmetric, but this non-axisymmetry is a remnant of the
initial, non-rotating triaxiality and is not due to the bar. We
measured the angle of the main halo part using the region with $10 > r
> 30$ kpc, thus avoiding the halo bar and the region outside it,
where it is not easy to disentangle 
the halo bar from the outer triaxiality, as well as the outermost
region where the density is low and which interacts little with the
disc component. We proceeded as discussed in Sect.~\ref{subsec:analysis} 
and plot the angle of the halo major axis in Fig.~\ref{fig:halovphi_angle}. 

We note that in all cases the haloes acquire some bulk rotation, even 
though the initial halo was built in equilibrium as non-rotating. 
In general, we find larger rotations 
for initially strongly non-axisymmetric haloes and for low
initial gas fractions. From Fig.~\ref{fig:A2Jhalo} we see that in
these cases more angular momentum is
given to the halo. It is therefore reasonable to assume that part of
the angular momentum absorbed by the halo is taken by the bulk
rotation and part by the motion of the individual `disc-like' halo
particles (see Sect.~\ref{subsec:halokinematics}). In good
agreement with this assumption, the halo rotates anticlockwise, i.e. in
the direct sense. 

We also note from Fig.~\ref{fig:halovphi_angle} that the angle of the
halo major axis also displays some
short period oscillations which we will discuss further in Sect.
\ref{sec:oscillations}. Mentally ignoring them, we see that 
the halo bulk rotation increases considerably with time, contrary to
the bar pattern speed, which has been shown to decrease with time  
(e.g. \citealt{Debattista.Sellwood.00}, A03, \citealt{Berentzen.SMVH.07}).

Fig.~\ref{fig:halovphi_angle} also shows that the halo bulk rotation
is very small, of the order of 5$^{\circ}$ -- 30$^{\circ}$ over a
period of 10 Gyr, i.e. even slower than
what was found by \cite{Bailin.Steinmetz.04}, \cite{Heller.SA.07} and
MA10, where the halo was found to rotate about 90$^{\circ}$ in a
Hubble time. The difference, however, is smaller 
if one takes into account the increase of the halo
bulk rotation with time and compares only the later times.

\section{Interaction between the various non-axisymmetric components}
\label{sec:oscillations}

In a simple barred galaxy, the bar is the only
non-axisymmetric component and its building blocks are
periodic orbits elongated along the bar
\citep{Contopoulos.Papayannopoulos.80, Athanassoula.BMP.83}. In the
case of a galaxy with two 
bars, an outer main bar and an inner small one, the basic building
blocks are loops, i.e. one-dimensional closed 
curves such that particles along them at a given time return to the same
curve (as viewed in the frame co-rotating with one of the bars) after
the two bars return to the same relative position
\citep{Maciejewski.Sparke.97}. It is intuitive that, in such cases,
the two bars can not rotate rigidly through each other, but should
show oscillations with the relative frequency of the two bars. This can
indeed be shown using the loop concept. Figure 2 in
\cite{Maciejewski.Sparke.00} and figure 6 in
\cite{Maciejewski.Athanassoula.07} show that a loop corresponding
to an inner bar is more elongated when the two bars are perpendicular
to each other and less elongated when the two bars are aligned. 

Double bars and the associated oscillations were also witnessed in a
number of simulations \citep[e.g.][and references below and
  therein]{Rautiainen.SL.02, Heller.SA.07L}. There is a general
agreement that the oscillation frequency is equal to the relative
frequency of the two bars, but considerable disagreement concerning the
remaining results. 
The simulations of \cite{Heller.SE.01} showed the formation of an inner 
ring/bar component which is more elongated when it is parallel to the
main bar, contrary to the loop predictions. A similar behaviour was
found for the double bar systems of \cite{Heller.SA.07}. Both these
works included gas in the simulations. On the other hand, 
\cite{Debattista.Shen.07} found that the strength of the outer bar has
a maximum when the two 
bars are aligned and a minimum when the the two bars are perpendicular
to each other. The strength of the inner bar is
maximum when the two bars are perpendicular and minimum when they are
aligned, in good agreement with the loop results
\citep{Maciejewski.Sparke.00, Maciejewski.Athanassoula.07} and the
simulations of \cite{Rautiainen.SL.02}. However these simulations and loop
calculations do not include gas. It is thus
of interest to revisit this question with our simulations.

\begin{figure}
\centering
\includegraphics[scale=0.7, angle=-90]{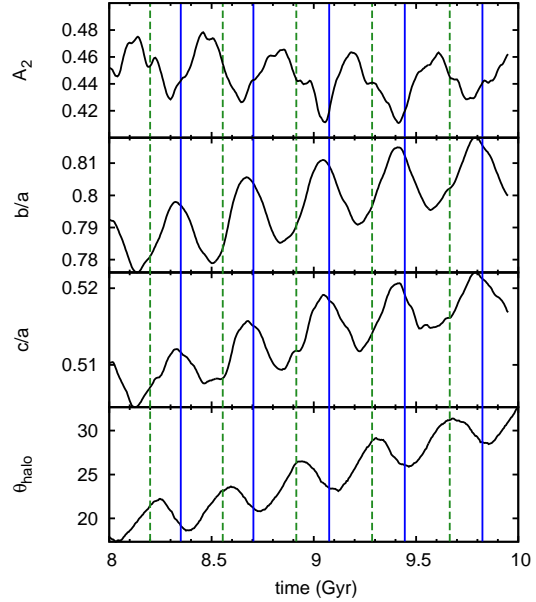}
\caption{Bar strength (upper panel), halo $b/a$ (second panel) and
  $c/a$ (third panel) axial ratios and angle of the halo major axis
  ($\theta_{halo}$), all as a function of time for a run with 0\% gas and an
  initially strongly triaxial halo (halo 3). The scales and the time 
  range were chosen so as to show best the oscillations. Vertical solid
  lines (blue in the online version) correspond to times when the
  directions of the halo and 
  the bar major axes are aligned, while dashed lines (green in the
  online version) correspond to the times when they are
  perpendicular. 
}
\label{fig:gtr003-t8-t10}
\end{figure}

\begin{figure}
\centering
\includegraphics[scale=0.7, angle=-90]{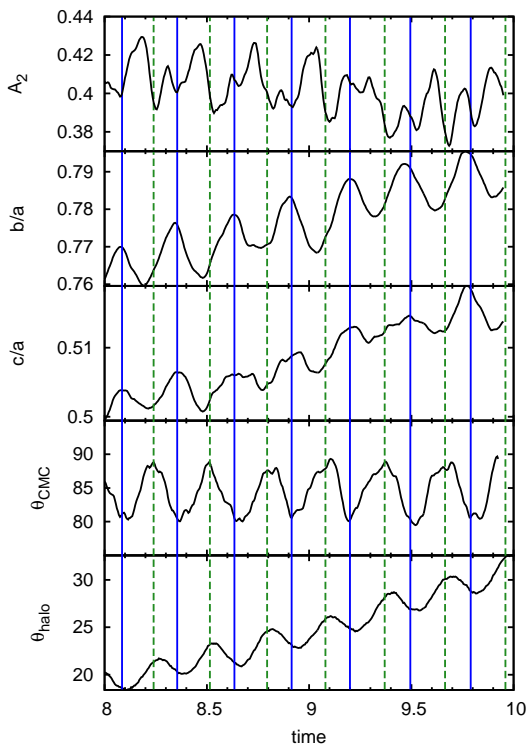}
\caption{Time evolution of the bar strength (upper panel), the halo
  $b/a$ and $c/a$ axial ratios (second and third panels from the top),
  the angle between the major axes of the gaseous CMC and the bar  
  (fourth panel from the top) and the angle for the halo major axis
  (lowest panel). for a simulation with
  initially 20\% gas and a strongly triaxial halo (halo 3). The scales
  and the time 
  range displayed are such so as to show best the oscillations. Solid
  lines (blue in the online version) correspond to times when the
  directions of the halo and the bar major axes are aligned, while the
  dashed lines (green in the online version) correspond to the
  times when they are perpendicular. 
}
\label{fig:gtr110-t8-t10}
\end{figure}

Figs.~\ref{fig:gtr003-t8-t10} and~\ref{fig:gtr110-t8-t10} show, for
two different simulations, the oscillations with time of the bar 
strength, of the halo axial ratios $b/a$ and $c/a$, and of the
angles of the halo and gas CMC major axes. The $b/a$ and $c/a$ were
calculated from the region with $10 > r > 30$ kpc, for the reasons
discussed in Sect.~\ref{subsec:halobulkrot}. We also indicated by
vertical lines in both figures the times at which the bar and the halo
major axes are parallel, or perpendicular. Note that, to zeroth
order approximation, when the bar and
halo major axes are aligned, the $A_2$ has a minimum and the $b/a$ has
a maximum, i.e. both are nearer to axisymmetry at 
that time. On the contrary, when the bar and halo major axes are
perpendicular, the $A_2$ has a maximum the $b/a$ has a minimum,
i.e. both are further away from axisymmetry at that time. The period of these
oscillations is compatible with the relative frequency of the bar and
halo rotation. Note also that the $c/a$ and the angles of the halo and
gaseous CMC major axes also have oscillations at the same
frequency. Those of $c/a$ are 
in phase with respect to the oscillations of $b/a$, while those of the
halo and gaseous CMC major axes are at a maximum when those of $b/a$ are at
a minimum. In as
much as comparisons are possible, the above results are in agreement with
those of the loop theory. 

Note, however, that this description is only a zeroth order
approximation of the simulation results, since
in fact the quantities whose time evolution we follow display patterns
which are more complex than single oscillations with a constant
frequency. This is due to  the self-consistency and
the strong non-linearity present in the simulations, which leads to 
results more complex than the simple models with rigid components and
constant pattern speeds can describe. For example, the
halo can not be described as a single rotating ellipsoid because the 
elongation in its inner parts follows the bar while the elongation of
its outer parts hardly rotates. More precisely, we should talk about two
groups of such components. On the one hand there is the triaxial halo,
which rotates with a very small angular velocity. On the other hand the main
bar, the halo bar and the gaseous and stellar CMCs, which are
non-axisymmetric components that rotate
with the bar pattern speed. Nevertheless, it is very gratifying that at
zeroth order approximations (i.e. as far as the simplifications
inherit in the loop theory can permit) there is no disagreement with known
theoretical results. 

The amplitude of these oscillations varies from one case to
another. It increases with increasing bar strength and also with
increasing halo triaxiality, as could be expected since strongest
interactions are expected when the two interacting non-axisymmetries
are strongest.

\section{Further discussion}
\label{sec:discussion}

\subsection{Advantages and limitations of this work}
\label{subsec:adv-lim}

Our simulations are dynamical rather than cosmological, i.e. they are
set so as to allow us to study best a given effect, in our case
the effect of gas and of non-axisymmetric haloes on the growth,
evolution and properties of the bar. Thus, as in all dynamical
simulations, the initial conditions are what is sometimes referred to as
idealised, i.e. the disc is assumed to have formed first and
simulations are started in the time interval after the disc has formed
and before the bar starts. This provides optimum conditions for
studies of bar formation and evolution. 

Compared to previous dynamical simulations, the ones presented here
have a number of strong 
points. The halo is live and is represented by one million particles,
a number which, for the adopted halo radial density
profile, is sufficient for an adequate description of the resonances
and therefore of the angular momentum exchange, thus not biasing the
whole evolution \citep[][A03]{Athanassoula.02}. We have also used a 
large number of gas particles, in all our standard cases with a mass of 
$m_{gas}=5 \times
10^{4}~M_{\odot}$ per particle. We ran also simulations with a much
higher number of gas particles, with a resolution up to $m_{gas}=2.5
\times 10^{3}~M_{\odot}$, 
and made sure there were no qualitative, or important quantitative 
differences. We, furthermore, have a high spatial resolution with a
gravitational softening of 50 pc. 

Contrary to most previous dynamical studies of bar formation and
evolution, our gas has both a cold and a hot phase and is modelled
including star formation, feedback and  
cooling. We do not claim that our recipes are perfect representations
of the interstellar medium. Indeed such perfect recipes are not
available \citep{Scannapieco.P.12}. But they are realistic, certainly
much more so 
than a complete neglect of star formation, which leads to a gas
fraction which does not decrease with time. The latter would entail
too low a gas fraction during the bar formation and early evolution
stages and/or a too 
high fraction during the secular evolution phase. A more in depth
description of the effect of various star formation, feedback and
cooling recipes on bar formation and evolution will be given elsewhere. 

We put considerable effort so that the initial
conditions we generate are as close to equilibrium as possible  
\citep{Rodionov.AS.09, Rodionov.Athanassoula.11}, 
so as to make sure that there are no transients due to inadequacies.
This allowed us, for example, to get information on how long the disc
can stay axisymmetric before forming the bar, or to measure the bar
growth rate. 

We made, whenever possible, comparisons between our results and
those of previous studies. In most cases, however, it was
not possible to make any quantitative comparisons, because in our
simulations the gas fraction varies with time, due to the star
formation, while in simulations with no star formation it stays
constant. It is thus unclear to what (constant) gas fraction value we
should be comparing our results to. Indeed, gas fraction can have
different effects on the different phases of the bar formation and
evolution.   

As limitations of this work we can say that we have considered 
only one mass model and only one set of star formation,
feedback and cooling recipes. Furthermore, we have not
included modules for chemical evolution, or for accretion, which would
have allowed us to follow jointly the chemical and the dynamical
evolution, nor have we discussed 
star formation, stellar populations and 
chemical abundances. These will be considered in future work. 

\subsection{The effect of gas on bar growth and evolution}
\label{subsec:gas-discuss}

\begin{figure}
\centering
\includegraphics[scale=0.39, angle=-90]{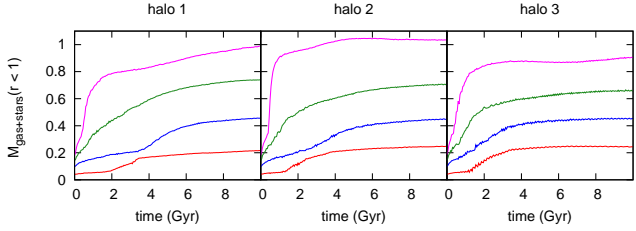}
\caption{Time evolution of the mass in the central region. We plot
  here the mass of the gaseous component and of the stars born from it,
  measured within the  central 1 kpc in units of $10^{10}
  M_{\odot}$. The three columns correspond to 
  the three different halo models. From left to right these are
  initially spherical, mildly triaxial and strongly triaxial. In the
  online version, the different
  initial gas fractions are shown with lines of different 
  colour, namely red for 20\%, blue for 50\%, green for
  75\% and magenta for 100\% gas.
}
\label{fig:CMCmass}
\end{figure}

It is clear from the previous sections that gas has a strong influence
on bar growth and evolution. Yet the relation is rather
complex. Indeed there are a number of effects, sometimes opposing each
other.

i) It is known that cold components respond more strongly to any forcing than
hot ones \citep{Binney.Tremaine.08}. Since gas is very cold and the
stars born from it are 
much colder than the old stars, one could expect that stronger
bars would be formed in more gas-rich cases.

ii) The gas also takes part in the angular momentum redistribution within
the galaxy. For gas-less simulations, frequency analysis of the orbits
has shown that angular momentum is emitted
from stars at (near-)resonance in the bar region and absorbed by
(near-)resonant material in the halo and in the outer disc. Thus the
angular momentum of the disc decreases with time and that of the halo
increases by an equal amount. If there is also a bulge, then it also can
absorb angular momentum. The above results were first found
and tested in simulations by Athanassoula (\citeyear{Athanassoula.02,
  Athanassoula.03} and unpublished), and then 
confirmed by a number of other simulations, with different models and
methods \citep[e.g.][]{Martinez.VSH.06, Ceverino.Klypin.07, Dubinski.BS.09,
Wozniak.Dansac.09, Saha.MVG.12}. 

Although frequency analyses, as the ones mentioned above, have not been 
yet performed for simulations including gas, both results from
several previous papers \citep[e.g.][]{Berentzen.AHF.04,
  Bournaud.CS.05, Berentzen.SMVH.07} and from our simulations  
argue that gas gives angular momentum to the bar and thus hinders its
growth. Thus the bar strength should decrease as the amount of gas
increases.    

iii) As already discussed in Sect.~\ref{subsec:longshortlived-context},
both orbital structure theory and N-body simulations have shown that
the existence of a massive and compact CMC can weaken the bar
strength. Contrary to some previous studies, in our simulations such a
CMC is not put in ``by hand'', it results from the evolution (Sect. 
\ref{subsec:CMClate}. Indeed, the bar pushes the gas inwards
where it forms a CMC. In Fig.~\ref{fig:CMCmass} we plot the mass of the gas
and of the \textsc{stars} in the inner one kpc as a function of time. 
It shows that in gas-poor simulations the CMC is 
considerably less massive than in the gas-rich ones. Thus, the CMC
will weaken the bar more in gas-rich simulations.       

iv) Stars are born in a very thin disc, which gradually thickens with time. 
The effect of the disc thickness on the bar strength and pattern speed
was studied  
by \cite{Klypin.VCQ.09}. They find that thinner discs host shorter and
weaker bars than thicker ones. Applying straightforwardly their result
to our simulations we can imply that simulations with less old stars 
and more gas and young stars will have a larger fraction
of their mass in a thin component, and therefore will give rise to
bars which are shorter and weaker than those which form in more
gas-poor discs.

We have thus listed four ways in which gas can influence the length
and strength of bars. Of these, three argue that more gas-rich galaxies
will form weaker bars, while one -- perhaps the least convincing one
-- argues for the opposite. In our simulations it is clear that bars
in gas-rich simulations are indeed weaker, arguing that the combined
effect of the angular momentum redistribution, of the CMC and of the 
vertical gas thickness dominates. 

\subsection{The structure of the CMC}
\label{subsec:CMC-discuss}

\subsubsection{The mass of the CMC as a function of gar fraction}
\label{subsubsection:CMCmass}

Fig.~\ref{fig:CMCmass} shows that the CMC mass included within one kpc
from the centre increases considerably with the gas fraction, and that
this holds for all the halo shapes we considered here. Interpreting this
is relatively  
complex, because there are three competing effects. 

i) The first has to do
with the total amount of gas in the simulation, since in
cases where this is very large, more gas will be pushed inwards, all
other quantities staying equal. 

ii) On the other hand, runs with a large
gas fraction have weaker bars (Sect.~\ref{subsec:allviews} and
~\ref{sec:barstrength}), so, all other quantities staying the 
same, less gas is pushed inwards. 

iii) A third, most important effect is
that the extent (both in radius and in energy range) of the $x_2$
orbits diminishes as the  bar strength  
increases \citep{Athanassoula.92a}. Therefore, and seen the
morphological results in Sect.~\ref{subsec:allviews}, one expects that the
$x_2$ is stronger in cases with more gas. 

Our results show that the
first and third effects coupled together are stronger than the second
one, so that the CMC component is more massive in simulations with 
more gas and less massive in simulations with less gas.

\subsubsection{The existence of both an inner and an outer Lindblad
  resonance}
\label{subsubsec:inner-outer-ILR}

In Sect.~\ref{subsec:CMClate} we showed that the gaseous CMC has two distinct
components. The first one, which we call the inner CMC, has a very small
extent and a very high density and it is elongated roughly along the
bar. The second one, which we call the outer CMC, has a considerably
larger extent, a lower (but still high) density and it is elongated
roughly perpendicular to the bar. The two together form the gaseous CMC.

To understand better these structures we froze the potential for a
number of simulations and times and calculated in each of those cases
a large number of orbits, using as initial conditions
the positions and velocities of simulation particles describing the
gas, or the very young stars, so as to follow the families of periodic
orbits \citep{Sanders.Huntley.76, Athanassoula.92a, Athanassoula.92b}. 
We followed them for 0.5 Gyr each, i.e. 
sufficiently so as to clearly get the orbital shape.
Amongst these orbits, we can divide the regular-looking ones into
three categories:

i) Orbits aligned roughly along the bar whose extent is of the order of the bar
length, or a sizeable fraction of it; 

ii) orbits aligned roughly along the
bar but whose extent is very much shorter than the bar, namely of the order of
the inner CMC or even less; 

iii) and orbits aligned roughly perpendicular
to the bar and whose extent is of the order of the outer CMC size or
even less. 

Orbital structure theory has shown \citep[e.g.][]{Contopoulos.80} that
the orientation of the 2:1 periodic orbits, in simple rigid potentials with
an axisymmetric and an $m$ = 2 part, changes by 90$^\circ$ at each
resonance. More specifically, it is parallel to the bar within the inner
ILR ($x_1$ family), perpendicular to it between the inner and the
outer ILRs ($x_2$ and $x_3$ families) and again parallel to it between
the outer ILR and the CR ($x_1$ family). Our orbital 
calculations confirm this for our more complex potentials and thus
explain the formation of the double CMC that we find.  

Note that such a CMC morphology can not be seen in the old stars. This
can be easily understood, because the velocity dispersion of old stars
is much larger than that of young stars \citep[e.g.][]{Nordstrom.P.04}.
Since the surface occupied by the $x_2$ family on a surface of section
is much smaller for larger energy values, it is much easier for gas
and very young stars to be trapped by the $x_2$ family than for old
stars. A fuller discussion of this has been given by
\cite{Patsis.Athanassoula.00}. 

It should be stressed that the existence of two ILRs, the inner and the
outer, hinges on the potential having an inner core and can never 
be found in potentials with a strong cusp. On the other hand, not any
potential with a core will have the gaseous CMC morphology that we
discussed hare, since this depends on the size of the core and the
strength of the bar. These will also influence very considerably the 
values of the sizes and size ratios for both components, since they
influence the extent and strength of the $x_2$ family.

This interpretation of the CMC structure 
makes a clear prediction. Indeed, orbital structure theory shows that
the maximum extent of the $x_2$ orbits  
is smaller than or of the order of the extent of the bar minor axis 
\citep[][figure 6]{Athanassoula.92a}. Thus, if
the outer component of the CMC is linked to the $x_2$ family, its
extent also should not exceed the size of the bar minor axis. This is not
as easy to test as in the models of \cite{Athanassoula.92a} because
the bar isodensities are not simple ellipses. To test it we made
images of the gas density in the inner part of the disc
on which we superposed the isodensities of the combined \textsc{disk}
and \textsc{stars} components (see the left panel of Fig.~\ref{fig:gasCMC}
for an example) and we created animations. By 
viewing both the individual images and the animations we were able to
verify that our prediction is indeed born out.
  
\subsection{Bar longevity}
\label{subsec:disc-longevity}

In our simulations bars never dissolved, even in cases with a very
large fraction of gas. This is an important element in the debate 
on whether bars are long- or short-lived, because our simulations have
live haloes, variable time-steps and a sufficient number of particles
in all components to describe the evolution adequately. They thus should
be giving the definitive answer, at least for the model we consider
here. Furthermore, those with an initially large fraction of gas mass 
also form quite massive CMCs. As can be seen in
Fig.~\ref{fig:CMCmass}, the CMC mass within a radius of one kpc from the
centre can reach between 4 and 20 per cent of the total disc mass,
depending on the total gas fraction. This is
very considerable, but still does not suffice to dissolve 
the bar because the CMC extent is relatively large. Indeed -- as has been
shown both by orbital structure work \citep{Hasan.PN.93} and by
simulations \citep[e.g.][]{Shen.Sellwood.04, Athanassoula.LD.05} -- for
a given mass, the more compact the CMC, the more it will reduce the bar
strength. If the same mass was concentrated within 100 pc rather than
one kpc, it would dissolve the bar, but neither observed CMCs, nor 
those grown self-consistently in simulations are that compact.

Several other simulations with different initial condition models
corroborate the fact that bars are long-lived. \cite{Villa.VSH.10} use
the same mass model as used here, but \cite{Berentzen.SMVH.07} have
different initial mass models. They also have a different description
of the gas, namely with an isothermal
equation of state with a temperature of $10^4$ K, or, in a few cases, with an
adiabatic equation of state. One of us (EA) ran a series of
simulations using a halo with a cusp, rather than a core as here, and
still found long-lived bars. Moreover, bars are also found to be long-lived in 
the cosmological simulations of e.g. \cite[][and in prep.]{Curir.MM.06,
  Scannapieco.Athanassoula.12} and particularly those of
\cite{Kraljic.BM.12}, again with different mass models and different
gas descriptions. 

Thus, the evidence in favour of bars being long-lived in isolated
galaxies is overwhelming. Could there, nevertheless, be exceptional
cases where the bars could dissolve? Since the two studies in which
bars did dissolve had rigid haloes, i.e. haloes which did not
participate in the angular momentum redistribution within the galaxy,
it makes sense to search in this direction, i.e. to make a simulation
with a rigid halo in which the bar dissolved and repeat it identical
except with a live halo, to see whether the bar still dissolves. A
further clue can come from the fact that, as we showed in
Sect.~\ref{sec:barstrength} that more flattened haloes inhibit bar
growth during the secular evolution phase. Thus, the ideal candidate
halo for allowing bar destruction would be a halo more squashed that 
$c/a$=0.4, i.e. disc-like, e.g. as advocated by
\cite{Pfenniger.CM.94}. The physical relevance of such haloes,
however, is not clear.

A more natural way of achieving bar destruction is via
interactions. Indeed, several simulations \citep{Pfenniger.91, 
Athanassoula.99, Berentzen.AHF.03} have shown that when a satellite
galaxy falls in a 
disc galaxy, it can destroy a pre-existing bar, while the disc
survives. This necessitates that the intruder is
sufficiently dense to reach the central regions while its mass is still
sufficiently high and also sets severe constraints on the geometry of
the encounter. Bar destruction in such mergings can be easily
understood because of 
the out-of-phase gravitational force the satellite exerts on the particles in
the bar, preventing them from following the $x_1$ family of orbits,
which is the backbone of the bar.    

\subsection{Comparison with observations. I}
\label{subsec:observationsI}

Figs.~\ref{fig:fig_xy_disk_6} and \ref{fig:fig_xy_disk_10} show that,
even for a single mass model, there is a variety of possible
morphologies of the bar and its surrounding region. Comparing with
images of observed galaxies, we see that all these morphologies are
realistic, so that it is not possible to put any observational
constraints from morphology alone. On the other hand, from the
measurements of bar strength discussed in Sect.~\ref{sec:barstrength}
and simple eye estimates of the bar length\footnote{For the present
  argument, simple eye estimates of the bar length are amply
  sufficient. More accurate measurements, using 
  other methods, will be given elsewhere. Comparison between various
  methods of measuring bar lengths can be found in
  \cite{Athanassoula.Misiriotis.02}, \cite{MichelDansac.Wozniak.06} and
  \cite{Gadotti.ACBSR.07}.}  from
Figs.~\ref{fig:fig_xy_disk_6} and \ref{fig:fig_xy_disk_10}, or from the
animations\footnote{\href{http://195.221.212.246:4780/dynam/movie/gtr}{http://195.221.212.246:4780/dynam/movie/gtr} }, we see that
our simulations give a very wide range of values, depending on the
time, the gas fraction and the halo shape. Observations show a similar
wide range of values, from the very strong bars such as NGC 4608 and
5701 discussed e.g. by \cite{Gadotti.deSouza.03} and shown to be of
the same strength as the bars in gas-less simulations, to the short
and/or weak bars in SAB types. 

In general, bars in gas-less simulations at times of the order of 10
Gyr or more, should have exceedingly strong
bars. Indeed, both cosmological simulations \citep{Kraljic.BM.12} and
observations \citep{Sheth.P.08} argue
that bars in very massive disc galaxies should be in place about 7 or 8 Gyr
back, while bars in lower-mass, blue spirals should be in place later,
perhaps as recently as 4 or 5 Gyr ago. Applying these numbers here
is not straightforward, because our simulations are dynamical and
start only when the disc 
is fully formed, but, even so, it is clear that 10 Gyr is an
overkill. Furthermore, discs do not form directly from stars. It is
gas that rains in from the halo on to the disc 
where it forms stars. Thus, even galaxies whose discs at the
present have very little gas, will have had much more at the time the
bar started forming, so that studying bar formation in gas-less simulations is
a further overkill. Therefore, bars in gas-less simulations at times of the
order of 10 Gyr or more should be very long and strong, much more so than
observed, so that when some gas is added and when shorter comparison
times are considered, the bars become realistic. Thus the mass model we propose
here fares well. 

\subsection{Comparison with observations. II}
\label{subsec:observationsII}

Our simulations help shed some light on the ``downsizing'' linked to
bar formation. As already mentioned above, bars in massive, red disc
galaxies are in place earlier than in blue, lower-mass spirals and the
time difference is important, of the order of 1 to 3 Gyr, or more
\citep{Sheth.P.08, Kraljic.BM.12}. \cite{Sheth.P.08}
tentatively brought up a possible explanation, namely that a low
mass disc is more harassed by a given perturber than a very heavy
one and that due to this harassment its bar can be destroyed. This
holds clearly for the case of mergers, but not necessarily for 
the case of interactions, which can, on the contrary, drive rather
than damp bars \citep{Gerin.CA.90, Miwa.Noguchi.98,
  Berentzen.AHF.04}. We will here propose an alternative explanation, but 
before discussing our alternative explanation, let us recall that 
observations show clearly that, both around redshift $z$
= 0 and at intermediate redshifts, smaller galaxies have a larger
fraction of gas than more massive ones \citep{Erb.SSPRA.06,
  Daddi.MCMWQ.10, TacconiP.10, Conselice.MBG.12}. 

Our explanation does not rely on interactions, but on the effect of
gas. As we saw in Sect. ~\ref{sec:barstrength}, 
bars form faster in gas-less or gas-poor simulations than in gas-rich
ones. This is due both to the fact that the initial simulation time during
which the disc stays roughly axisymmetric is shorter (i.e. the bar
starts growing earlier) and to the fact that, once it has started
growing, the bar grows much faster. The two together argue strongly
that bars 
will be in place in red galaxies much before they are in blue ones. 
Comparing two simulations identical in everything except for the
gas-to-total mass ratio in the disc, we find that the difference
between their bar formation times is in good agreement 
with what observations show us. Comparing e.g. simulations
101 and 111 we find that the former reaches a bar strength  $A_2$ =
0.3 (0.4) before the latter with a time difference 2.1 (2.9) Gyr. 
This not only explains the delay, but also gives an estimate of the
delay time which is in good agreement with observations.
It can explain why bars are in place at earlier times in massive
galaxies and at later times in blue, less massive ones.
Of course, there are many differences between these two types of
galaxies, other than the fraction of gas in the disc. These can include
the total mass and extent of the galaxy, the form of its
rotation curve and the existence of a bulge component. Nevertheless,
our simulations argue that the effect of the gas can, by itself, go a
long way towards accounting for the difference between bar formation
times of red, massive galaxies and of blue, lower-mass ones.

\subsection{Comparison with observations. III}
\label{subsec:observationsIII}

Our work also introduces a number of further possibilities for confrontation
with observations. For example, both a qualitative and a quantitative 
comparison of the new morphological features, discussed in the end of
Sect.~\ref{subsec:allviews} and Sect.~\ref{subsec:CMC-discuss}, to
analogous components in real galaxies would be very
useful. Furthermore,  
thorough comparisons of the morphology of the
observed gaseous CMC to our results in Sect.~\ref{subsec:CMClate}
should now be feasible with ALMA. In particular, it would be
interesting to search for the existence of structures similar to those
of the inner gaseous CMC. Indeed, as already mentioned, such a
structure would only exist in galaxies with a core, so that the
existence of such a structure would give further observational
evidence for the existence of inner cores in galaxies and thus further
input to the core versus cusp debate. On the other hand, the lack of
such a structure would not necessarily point to a cusp. 

A further, very worthwhile project is to study the kinematics of the
halo stars located in the vicinity of the Galactic disc and this should be
possible by comparing results from Sect.~\ref{subsec:halokinematics}
with measurements from GAIA and from GAIA-related 
kinematical surveys on large ground-based telescopes.
As already discussed, simulations show that the halo has a quite
complex structure with in 
its inner part either a halo bar, or a close-to-axisymmetric region,
depending on the bar strength. At larger distances from the centre the
halo can stay triaxial. Analysis of the
kinematics of a sufficient number of disc and halo stars of our Galaxy
-- as will be available from GAIA and from the related
kinematical surveys on large ground-based telescopes -- should shed
light and set strong
constraints on any theoretical study of the local disc/halo
interaction. Inversely, any such 
theoretical study should allow us to explain and model these GAIA
data.

\section{Summary and conclusions}
\label{sec:conclusions}

In this paper we discussed the evolution of barred galaxies using
simulations including gas and/or an initially triaxial halo. 
We showed that both the gas and the halo triaxiality influence
strongly the bar formation, evolution and properties. In turn, the bar
influences their properties and dynamics of its host galaxy, such as
the gas surface density distribution, 
the halo shape and kinematics, as well as the redistribution of
angular momentum within the galaxy.

In our simulations the gas fraction decreases with time due to star
formation. Starting from a ratio of gas to total disc mass covering
the whole range 
between 0\% and 100\%, it comes to levels well compatible with the
fraction of gas in disc galaxies in the nearby universe, and at
intermediate redshifts. Since our initial conditions were set up so as
to be as near equilibrium as possible, in cases with initially
strongly triaxial haloes both the \textsc{disk} and 
the gas components start off non-axisymmetric, although less so than
the halo. This lasts for at least one Gyr. During this time, the disc
of young stars forms from 
the centre outwards, as expected, and both it and the gas show a clear
multi-arm spiral structure.  

\subsection{Gas}
\label{subsec: conclusions-gas}

In gas-rich discs, the disc stays near-axisymmetric much
longer than in gas-poor cases, and, when the bar starts growing it
does so at a much slower rate (Fig.~\ref{fig:A2-boa-coa}). These two
results, taken together, can 
explain the observation that bars are in place earlier in massive red
disc galaxies than in blue spirals \citep{Sheth.P.08}. We also find that the
morphological characteristics of both the gaseous and the stellar
density distribution in the bar region are strongly influenced 
by the gas fraction.

A further important result is that the bar in gas-poor or gas-less
cases grows to become longer and reaches a higher amplitude
than in gas-rich cases, provided that the remaining
parameter values are the same. We wish to stress, however, that
increasing the gas fraction is not the only way to obtain this
effect. Indeed, as already discussed in A03 for gas-less simulations,
the length and strength of the
bar are directly related to the angular momentum redistributed within
the galaxy. The latter is influenced by the amount of angular momentum
the halo can absorb, by the existence (or absence) of a 
classical bulge and its radial extent as well as the velocity
anisotropy of all components (A03). \cite{Klypin.VCQ.09} discussed
the effect of the disc thickness. Here we added the effects
of gas and halo shape, and their relative importance.   

The bar in our simulations pushed the gas inwards to form a central
mass concentration (CMC) with a very high gas density, resulting in
considerable star formation. The mass of this CMC is in the range
between 4 and 20 \% of the total disc mass and it is more massive 
in more gas-rich simulations. In cases with an initially purely 
gaseous disc, its
mass within the inner kpc can reach as high as $10^{10} M_{\odot}$.  
Thanks to the high resolution of our simulations, we were able to unravel
the complex morphology of the gaseous CMC. We identified two
components, a small inner one, elongated parallel to the bar, and a
considerably larger one elongated perpendicular to it. Calculating
orbits, we were able to show that this structure is due to a double
ILR, the inner CMC being within the inner ILR and the outer one
between the inner and the outer ILR. 

A question which has been considerably debated over the last ten years
or so, is whether bars are long-lived or short-lived. Indeed, a number
of previous works have given conflicting results, but all of them had at
least one caveat, concerning the number of gaseous particles, the
spatial resolution 
and/or the appropriate description of halo resonances. Our simulations
do not have either of these caveats and, furthermore, they have a
resolution of 50 pc. In these simulations, we never witnessed any
destruction or dissolution of a large scale bar, which strongly argues 
for long-lived bars, at least in isolated galaxies. 

\subsection{Halo}
\label{subsec: conclusions-halo}

The effect of the halo relative mass on bar evolution is dual, as
first discussed by \cite{Athanassoula.02}. Namely, during the early phases
of bar formation the halo delays bar formation, while during the later
secular evolution phase it makes the bar stronger. In this paper we
showed that the effect of the halo shape on bar evolution is also 
dual. Thus, in the initial part of the evolution the 
non-axisymmetric forcing of the triaxial halo can trigger bar
formation, so that bars in triaxial haloes grow earlier than in 
spherical ones. On the contrary, at the later stages of evolution,
when the bar is well grown, halo triaxiality hinders bar growth due to 
the nonlinear interaction between the two non-axisymmetries. Thus  in
galaxies with strongly triaxial haloes bars grow earlier, but
their strength hardly increases during the secular evolution phase.
  
We extend previous results on the halo bar (or dark matter bar) found
in the innermost parts of the halo of gas-less models 
\citep[e.g.][]{Athanassoula.05b, Athanassoula.07, Colin.VK.06} and we show
that it also exists in gas rich models with initially spherical haloes
and also in gas-less or gas-poor models with initially
triaxial haloes. The inner parts of models with no halo bar show, on
the contrary, an inner region which is more axisymmetric than the part
outside it. Our simulations show that a very rough limit between the
two types of models can be set 
by their bar strength, strong bars with $A_2>0.5$ leading to a halo
bar and weaker bars with $A_2<0.5$ leading to a halo inner part which is
more axisymmetric than further out.   

We also calculated the halo axial ratio in regions outside the halo
bar. Using the region 10 $> r >$ 30 kpc, we find that 
the time evolution of the halo axial ratios
is clearly linked to the time evolution of the bar
strength. The strongest evolution occurs at 
times between bar growth and secular evolution. 

There is a strong correlation between the bar strength and the amount of
angular momentum absorbed by the halo, as has already been found for gas-less
simulations with initially spherical haloes (A03).
We find, however, that haloes with strong initial triaxiality, which
stays strong during the simulation, deviate from this line following
their own regression. This is in agreement with our results on the
dual effect of the halo shape.

Part of the angular momentum absorbed by the  halo changes considerably 
the halo kinematics and particularly that of the material near
the disc equatorial plane, which acquires considerable rotation. This
rotation correlates well with the bar strength and we find a Pearson
correlation coefficient of the order of 0.9 between the maximum
tangential velocity and the $A_2$ measure of the bar strength. 

Another part of the angular momentum absorbed by the  halo provides
the halo with a bulk rotation, which is usually negligible over the
first few Gyr and then increases steadily with time, contrary to that
of the disc bar which decreases with time. Nevertheless, on average 
the position angle of the halo changes very little, of the order of
5$^{\circ}$ -- 30$^{\circ}$ 
over a period of 10 Gyr. In general, we find larger rotations 
for initially strongly non-axisymmetric haloes and for low
initial gas fractions (i.e. strong bars). 
 
In models with triaxial haloes, a number of the main quantities
describing the model -- such as the bar
strength, the halo axial ratios $b/a$ and $c/a$, the angle of the
major axis of the halo and of the gaseous CMC, etc -- show
clear oscillations in their time evolutions. These are due to the
presence of several 
non-axisymmetric components, and they do not all rotate
with the same pattern speed. 
We found that, to zeroth order approximation, when the bar and
halo major axes are aligned, the $A_2$ has a minimum and the $b/a$ has
a maximum, i.e. in both cases the $m$ = 2 components are less
strong. On the contrary, when the bar and halo major axes are 
perpendicular, the $A_2$ has a maximum and the $b/a$ has a minimum,
i.e. the corresponding $m$ = 2 components are 
stronger. Furthermore, the
period of these variations is compatible with the relative frequency
of the bar and halo rotation and the angles of the halo and of the
gaseous CMC are also locked in the same oscillatory pattern. 


\section*{Acknowledgements}

It is a pleasure to thank Volker Springel for making available to us
the version of \textsc{gadget2} used here, Albert Bosma and Arman Khalatyan
for useful discussions and Jean-Charles Lambert for the uns and GLNEMO
software (http://projets.oamp.fr/projects/glnemo2). 
EA acknowledges financial support to the DAGAL network from the People
Programme  (Marie Curie Actions) of the European Union's Seventh
Framework Programme FP7/2007-2013/ under REA grant agreement number
PITN-GA-2011-289313. EA also acknowledges financial support from the
CNES (Centre National d'Etudes Spatiales - France).               
REGM acknowledges support from the Brazilian agencies FAPESP
(05/04005-0) and CAPES (3981/07-0), and from the French Ministry of Foreign and
European Affairs (Eiffel fellowship). 
SR acknowledges support from the Russian Foundation for
Basic Research (grants 08-02-00361-a and 09-02-00968-a)
and a grant from the President of the Russian
Federation for support of Leading Scientific Schools (grant
NSh-1318.2008.2). 

\bibliographystyle{mn2e.bst}
\bibliography{rm_3axhalo_gas.bbl}


\label{lastpage}

\end{document}